\documentclass[
 aip,
 amsmath,amssymb,
reprint,%
]{revtex4-2}

\usepackage{graphicx}
\usepackage{color} 
\usepackage{dcolumn}
\usepackage{bm}

\usepackage[utf8]{inputenc}
\usepackage[T1]{fontenc}
\usepackage{mathptmx}
\usepackage[version=4]{mhchem}
\usepackage{siunitx}
\usepackage{amsfonts}
\usepackage{float}
\usepackage{longtable}
\usepackage{tabularx}
\usepackage{multirow}
\usepackage{rotating}

\usepackage{hyperref}
\hypersetup{
    colorlinks=true,
    linkcolor=blue,
    filecolor=magenta,      
    urlcolor=cyan,
}

\begin{document}

\preprint{AIP/123-QED}

\title{NMR Chemical Shift Computations at Second-Order Møller-Plesset Perturbation Theory
Using Gauge-Including Atomic Orbitals and Cholesky-Decomposed Two-Electron Integrals}

\author{Sophia Burger}
 \email{soburger@uni-mainz.de}
 \affiliation{Department Chemie, Johannes Gutenberg-Universit{\"a}t Mainz, Duesbergweg 10-14, D-55128 Mainz, Germany}
\author{Filippo Lipparini}%
 \email{filippo.lipparini@unipi.it}
\affiliation{Dipartimento di Chimica e Chimica Industriale, Universit\`{a} di Pisa, Via G. Moruzzi 13, I-56124 Pisa, Italy}
\author{J{\"u}rgen Gauss}%
 \email{gauss@uni-mainz.de}
\affiliation{Department Chemie, Johannes Gutenberg-Universit{\"a}t Mainz, Duesbergweg 10-14, D-55128 Mainz, Germany}
\author{Stella Stopkowicz}%
 \email{stella.stopkowicz@uni-mainz.de}
\affiliation{Department Chemie, Johannes Gutenberg-Universit{\"a}t Mainz, Duesbergweg 10-14, D-55128 Mainz, Germany}
\
%
\date{\today}

\begin{abstract}
We report on a formulation and implementation of a scheme to compute NMR shieldings at second-order M{\o}ller-Plesset (MP2) perturbation theory using gauge-including atomic orbitals (GIAOs) to ensure gauge-origin independence and Cholesky decomposition (CD) to handle unperturbed as well as perturbed two-electron integrals. We investigate the accuracy of the CD for the derivatives of the two-electron integrals with respect to an external magnetic field as well as for the computed NMR shieldings, before we illustrate the applicability of our CD based GIAO-MP2 scheme in calculations involving up to about one hundred atoms and more than one thousand basis functions.
\end{abstract}
\maketitle

\section{\label{intro}{Introduction}}
The computation of NMR chemical shifts is
an important application of quantum chemistry.\cite{Kaupp04}
For the accurate computation of NMR shieldings it has been amply shown that the consideration of electron-correlation effects is essential.\cite{Buehl93,Gauss95c,Stanton96a,Gauss02c,Harding08c,Harding11,Teale13,Jaszunski21}
In this respect, second-order Møller-Plesset (MP2) perturbation theory\cite{Moller34,Cremer00} has
 been shown to be very useful,\cite{Gauss92,Gauss93,Buehl93,Sieber93}
even though highly accurate predictions of NMR shieldings, 
in particular when aiming at absolute shieldings,\cite{Gauss95a,Gauss95b,Gauss96a,Sundholm96,Puzzarini09}
require
coupled-cluster (CC) treatments.\cite{Shavitt09}
Nevertheless, as the calculation of relative NMR chemical shifts benefits from some error cancellation, MP2 computations can provide a very useful and reliable tool. However, MP2 computations of
NMR shieldings are quite costly in comparison to corresponding
Hartree-Fock (HF) and density-functional theory (DFT) treatments and thus DFT
computations are currently the first choice for NMR shielding computations,
despite certain deficiencies of standard DFT in treating magnetic properties.\cite{Teale13} Efforts to speed up MP2 computations of NMR shieldings and thus to
increase the applicability of MP2 have a long history. Integral-direct schemes
 together with an efficient exploitation of point-group symmetry and coarse-grain
 parallelization have significantly extended the applicability of MP2.\cite{Kollwitz96,Kollwitz98}
The use of local-correlation treatments
has further enhanced the applicability of MP2, as shown by the work of Loibl and Schütz.\cite{Loibl12}
These authors also use density fitting\cite{Whitten73,Dunlap79,Eichkorn95} to avoid the computation of the perturbed two-electron integrals when using gauge-including
atomic orbitals (GIAOs).\cite{London37,Hameka58,Ditchfield72,Wolinski90,Helgaker91} Maurer and Ochsenfeld\cite{Maurer13} reported on a Laplace-based GIAO-MP2 formulation and implementation for the computation of NMR chemical shifts. This scheme, together with an efficient implementation, should in principle allow to achieve linear scaling.
Stoychev {\it  et al.}\cite{Stoychev21} recently described the implementation of a scheme to compute 
GIAO-MP2 NMR chemical shifts within the domain-based local pair natural orbital (DLPNO) framework. 
The applications reported by these authors involved cases with up to 4700 basis functions and convincingly demonstrate the efficiency of this implementation which is part of the ORCA package.\cite{Neese20} While the work by Stoychev {\it et al.} achieves computational efficiency by the use of a local-correlation treatment, of pair natural orbitals,\cite{Neese09} and of density fitting,\cite{Stoychev18a,Stoychev18} together with an efficient exploitation of sparsity,\cite{Pinski15} there are more possibilities to reduce the overall cost of the calculation. 
One option is here to apply a Cholesky decomposition (CD) to the two-electron integrals.\cite{Beebe77,Koch03} This option has in particular proven useful for medium-sized systems,\cite{Epifanovsky13,Nottoli21} as CD alone does not have the potential to reach linear scaling. The advantage of CD over density fitting is that no auxiliary basis sets are required and that the error can be controlled in a rigorous manner. While there is an extensive literature\cite{Koch03,Aquilante07a,Aquilante07b,Boman08,Aquilante09a,Aquilante09b,Epifanovsky13,Folkestad19} on the use of CD within energy computations, less work has been reported for property computations. Recently, CD has been applied to accelerate the computation of nuclear gradients,\cite{Bostroem14,Feng19} but nothing has been so far reported concerning the use of CD for the computation of magnetic properties.

In this paper, we describe how CD can be used for the perturbed two-electron integrals that appear in GIAO computations of NMR shieldings and demonstrate the efficiency of a corresponding CD treatment for GIAO-MP2 computations on medium-sized systems with more than one thousand basis functions. In the following, after a brief review of standard GIAO-MP2 theory and CD, we discuss how CD is applied to the perturbed two-electron integrals 
and describe how CD can be exploited in GIAO-MP2 computations. The theory section is followed by a description of our implementation within the CFOUR program package.\cite{cfour,Matthews20c} We then discuss the accuracy of CD in GIAO-MP2 computations, before demonstrating the computational efficiency of our CD based GIAO-MP2 scheme in calculations. 

\section{Theory}

\subsection{\label{sec2a}Standard GIAO-MP2 theory}
\label{theor1}
We start by recapitulating the standard theory for the computations of NMR shieldings at the MP2 level when using GIAOs. We follow here closely Ref.~\onlinecite{Kollwitz96} in which spin-adapted expressions have been given.

The NMR shielding tensor $\boldsymbol{\sigma}^N$ of the $N$th nucleus in a molecule is defined as the second derivative of the energy with respect to the external magnetic field $\bf B$ and the magnetic moment ${\bf m}_N$ of the $N$th nucleus and most conveniently evaluated using the following density-matrix based expression:
\begin{eqnarray}
\label{eq1}
\sigma_{ji}^{N} &=& \sum_{\mu \nu} D_{\mu \nu}\
{{\partial^2 h_{\mu \nu}} \over {\partial  B_i\ \partial  m_{Nj}}}
+ \sum_{\mu \nu} {{\partial D_{\mu \nu}} \over {\partial  B_i}}\
{{\partial h_{\mu \nu}} \over {\partial m_{Nj}}}.
\end{eqnarray}
In Eq.~(\ref{eq1}), $D_{\mu \nu}$ refers to the one-particle density matrix, with the Greek indices $\mu$ and $\nu$ labeling atomic orbitals (AOs), and $h_{\mu\nu}$ denotes the matrix elements of the one-electron Hamiltonian. Expression for the derivatives of $h_{\mu\nu}$ with respect to $B_i$ and/or $m_{Nj}$ can be, for example, found in Ref.~\onlinecite{Gauss93}. The MP2 contribution to the density matrix $D_{\mu\nu}$ is usually defined in the corresponding molecular-orbital (MO) representation
\begin{eqnarray}
\label{eq2}
D_{\mu \nu} &=& \sum_{pq} c^*_{\mu p}\ D_{pq} c_{\nu q},
\end{eqnarray}
with $c_{\mu p}$ specifying the MO coefficients obtained by solving the HF self-consistent-field (HF-SCF) equations. 
Indices $i,j, \dots$ label in the following occupied spatial orbitals, indices $a,b, \dots$ denote virtual spatial orbitals, and indices $p,q, \dots$ generic MOs that are either occupied or unoccupied. In the MO representation, the occupied-occupied and virtual-virtual block of the MP2 density matrix are 
given by
\begin{eqnarray}
\label{eq3}
D_{ij} &=&  - 2 \sum_m \sum_{ef} \tilde{t}_{im}^{ef}\ {t_{jm}^{ef}}^*
\end{eqnarray}
and
\begin{eqnarray}
\label{eq4}
D_{ab} &=& 2 \sum_{mn} \sum_e {\tilde{t}_{mn}^{ae}}{}^{*}\ t_{mn}^{be}.
\end{eqnarray}
Note that the given expression for the occupied-occupied block does not include the HF contribution to the
density matrix.
The MP2 amplitudes in Eqs.~(\ref{eq3}) and (\ref{eq4}) are defined as
\begin{eqnarray}
t_{ij}^{ab} = \frac{(ai|bj)}{\epsilon_i + \epsilon_j - \epsilon_a - \epsilon_b}
\end{eqnarray}
with $(pq|rs)$ as the MO two-electron integrals in Mulliken notation,
$\epsilon_p$ as the orbital energy of the $p$th orbital, and $\tilde{t}_{ij}^{ab}$ as the corresponding ``spin-adapted'' amplitudes
\begin{eqnarray}
\tilde{t}_{ij}^{ab} = 2 t_{ij}^{ab} - t_{ji}^{ab}.
\end{eqnarray}
The virtual-occupied block of the MP2 density matrix is obtained by solving the Z-vector equations\cite{Handy84}
\begin{eqnarray}
\label{eq5}
 \sum_m \sum_e  D_{em} [ 4 (em|ia) - (ea|im) - (ma|ie)
 \nonumber\\ + \delta_{im} \delta_{ae}
(\epsilon_a - \epsilon_i) ]
= - 2 X_{ai}
\end{eqnarray}
with the intermediate $X_{ai}$ given by
\begin{eqnarray}
\label{eq6}
X_{ai} & = & \sum_{m} \sum_{ef} (ea|fm)
{\tilde{t}_{im}^{ef}}{}^* - \sum_{mn} \sum_e (im|en)
{\tilde{t}_{mn}^{ae}}{}^*
+ \sum_{mn} D_{mn} \bigg\{(mn|ia) \nonumber\\ && \left.- \frac{1}{2} (ma|in)\right\} 
+ \sum_{ef} D_{ef} \left\{(ef|ia) - \frac{1}{2}
(ea|if)\right\} .
\end{eqnarray}

Expressions for the perturbed MP2 density matrix can be obtained by straightforward differentiation of Eqs.~(\ref{eq3}), (\ref{eq4}), (\ref{eq5}), and (\ref{eq6}) with respect to the components $B_i$ of the external magnetic field.
The corresponding expressions for the occupied-occupied and virtual-virtual block of the perturbed MP2 density matrix are
\begin{eqnarray}
\label{eq8}
{{\partial D_{ij}}\over {\partial B_i}}& =& -2 \sum_m \sum_{ef} \left\{
 {{\partial {t}_{im}^{ef}} \over {\partial B_i}} {\tilde{t}_{jm}^{ef}}{}^*\ 
+ {{\partial {t_{jm}^{ef}}^*} \over {\partial B_i}} \tilde{t}_{im}^{ef}
\right\}
\end{eqnarray}
and
\begin{eqnarray}
\label{eq9}
{{\partial D_{ab}}\over {\partial B_i}} &=&  2  \sum_{mn} \sum_e \left\{
 {{\partial {{t}_{mn}^{ae}}^*} \over {\partial B_i}} \tilde{t}_{mn}^{be}
+ {\tilde{t}_{mn}^{ae}}{}^* {{\partial t_{mn}^{be}} \over {\partial B_i}} \right\},
\end{eqnarray}
while the perturbed virtual-occupied block is obtained as the solution to the
perturbed Z-vector equation
\begin{eqnarray}
\label{eq10} 
\sum_m \sum_e {{\partial D_{em}} \over { \partial B_i}} \left\{ 
  (ma|ie) - (ea|im) + \delta_{im} \delta_{ea} (\epsilon_a - \epsilon_i) \right\}
\nonumber \\  = - 2 {{\partial X_{ai}}\over{\partial B_i}} 
- \sum_m \sum_e D_{em}
\left\{
2 {{\partial (em|ia)}\over{\partial B_i}}
+ 2 {{\partial (me|ia)}\over{\partial B_i}} \right.\qquad\qquad\nonumber \\  \left.
- {{\partial (ea|im)}\over{\partial B_i}}
- {{\partial (ma|ie)}\over {\partial B_i}}
+ \delta_{im} {{\partial f_{ea}}\over{\partial B_i}} - \delta_{ea}
{{\partial f_{im}}\over {\partial B_i}} \right\}\qquad\qquad
%
\end{eqnarray}
with $\partial X_{ai}/\partial B_i$ defined by
\begin{eqnarray}
\label{eq11}
{{\partial X_{ai}}\over {\partial B_i}} & = &
\sum_m \sum_{ef} \left\{ {{\partial {\tilde{t}_{im}^{ef}}{}^*} \over
{\partial B_i}} (ea|fm) + \tilde{t}_{im}^{ef}{}^* {{\partial
(ea|fm)}\over {\partial B_i}} \right\} 
\nonumber\\&& - \sum_{mn} \sum_e \left\{ {{\partial {\tilde{t}_{mn}^{ae}}{}^* }\over {\partial
B_i}} (im|en)  + {\tilde{t}_{mn}^{ae}}{}^* {{\partial
(im|en)
} \over {\partial B_i}} \right\} \nonumber\\&& 
+ \sum_{mn} \left\{ {{\partial D_{mn}}\over {\partial B_i}} \left[(mn|ia)
-
\frac{1}{2} (ma|in)]
\right] \right.  \nonumber \\ && \quad\quad\quad \left.
+ D_{mn}
\left[{{\partial (mn|ia) } \over {\partial B_i}}
- \frac{1}{2} {{\partial (ma|in) } \over {\partial B_i}}\right]
\right\}  \nonumber \\ &&
+ \sum_{ef} \left\{ {{\partial D_{ef}} \over {\partial B_i}} \left[
(ef|ia) - \frac{1}{2} (ea|if) \right] 
 \right.  \nonumber \\ &&\quad\quad\quad \left.
 + D_{ef} 
\left[{{\partial (ef|ia) } \over {\partial
B_i}} 
- \frac{1}{2} {{\partial (ea|if) } \over {\partial
B_i}}\right] \right\}.
\end{eqnarray}
In the given equations ${{\partial f_{pq}}\over {\partial B_i}}$ denotes the perturbed Fock matrix,
${\partial (pq|rs) } \over {\partial B_i}$ the perturbed integrals
\begin{eqnarray}
\label{eq12}
{{\partial (pq | rs ) } \over {\partial B_i}} & = &
+\sum_{\mu \nu \rho \sigma}
{{\partial {c_{\mu p}}^*}\over{\partial B_i}} c_{\nu q} {c_{\sigma r}}^*
c_{\rho s} {( \mu \nu | \sigma \rho)}\nonumber \\ &&
+\sum_{\mu \nu \rho \sigma} {c_{\mu p}}^* 
{{\partial c_{\nu q}}\over{\partial B_i}} {c_{\sigma r}}^*
c_{\rho s} {( \mu \nu | \sigma \rho)}
\nonumber \\ && 
+\sum_{\mu \nu \rho \sigma} {c_{\mu p}}^* c_{\nu q}
{{\partial {c_{\sigma r}}^*}\over{\partial B_i}}
c_{\rho s} {( \mu \nu | \sigma \rho)}\nonumber \\ &&
+\sum_{\mu \nu \rho \sigma} {c_{\mu p}}^* c_{\nu q} {c_{\sigma r}}^*
{{\partial c_{\rho s}}\over{\partial B_i}}
{( \mu \nu | \sigma \rho )}  \nonumber \\ &&
+\sum_{\mu \nu \rho \sigma} {c_{\mu p}}^* c_{\nu q} {c_{\sigma r}}^*
c_{\rho s} {{ \partial (\mu \nu | \sigma \rho) } \over {\partial
B_i}},
\end{eqnarray}
${{\partial t_{ij}^{ab}} \over {\partial B_i}}$ the perturbed amplitudes
\begin{eqnarray}
\label{eq13}
{{\partial t_{ij}^{ab}}\over {\partial B_i}} & = & \left\{
{{\partial (ai|bj)} \over {\partial B_i}}
+ \sum_e \left[ {{\partial f_{ae}} \over {\partial B_i}} t_{ij}^{eb}
+ {{\partial f_{be}} \over {\partial B_i}} t_{ij}^{ae} \right] \right. \nonumber \\ && \left.
- \sum_m \left[ t_{mj}^{ab} {{\partial f_{mi} }\over {\partial B_i}}
+ t_{im}^{ab} {{\partial f_{mj} }\over {\partial B_i}} \right]
\right\} / (\epsilon_i + \epsilon_j - \epsilon_a - \epsilon_b), \nonumber \\ &&
\end{eqnarray}
and $\partial c_{\mu p} \over {\partial B_i}$ the derivatives of the MO coefficients $c_{\mu p}$ with respect to the components of the magnetic field. The latter are determined via the coupled-perturbed HF (CPHF) equations.\cite{Gerrat68,Pople79}

Note that the non-vanishing HF contribution to the density matrix is given by
\begin{eqnarray}
D_{ij}^\mathrm{HF} = 2 \delta_{ij}
\end{eqnarray}
and that there is no HF contribution to the perturbed density matrix in the MO representation.

The AO representation of the perturbed density matrix is finally obtained as
\begin{eqnarray}
\label{eq14}
{{\partial D_{\mu \nu}} \over {\partial B_i}} &=&
\sum_{pq} \left\{ c_{\mu p}^* {{\partial D_{pq}} \over {\partial B_i}} c_{\nu q}
+ 
{{\partial c_{\mu p}^*}\over{\partial B_i}} D_{pq} c_{\nu q}
+
c_{\mu p}^* D_{pq} {{\partial c_{\nu q}}\over{\partial B_i}} \right\}.
\nonumber\\ &&
\end{eqnarray}

\subsection{\label{sec2b}Cholesky decomposition of two-electron integrals}

To reduce computational cost and in particular the memory requirements for handling the two-electron integrals, a Cholesky decomposition (CD)\cite{Beebe77,Koch03} can be applied to the positive semi-definite two-electron integral matrix\footnote{Note that we define the two-electron integral matrix with the indices for electron 2 interchanged in order to ensure positive semi-definiteness of the matrix in the presence of an external magnetic field, see Ref.~\onlinecite{Blaschke21}}
\begin{eqnarray}
\label{cd1}
(\sigma \rho | \nu \mu) \approx \sum_{P=1}^M L_{\sigma \rho}^P {L_{\mu\nu}^P}^*
\end{eqnarray}
with $M$ as the rank of the decomposition and the Cholesky vectors (CVs) $L_{\mu \nu}^P$ determined via
\begin{eqnarray}
\label{cd2}
L_{\sigma \rho}^P = \widetilde{(\mu \nu | \nu \mu)}^{- \frac{1}{2}}
\left\{(\sigma \rho | \nu \mu ) - \sum_{R=1}^{P-1} L_{\sigma \rho}^R {L_{\mu\nu}^R}^*\right\}
\end{eqnarray}
with the updated diagonal elements of the two-electron integral matrix given by
\begin{eqnarray}
\widetilde{(\mu \nu | \nu \mu)} = (\mu \nu | \nu\mu) - \sum_{R=1}^{P-1} L_{\mu \nu}^R {L_{\mu \nu}^R}^*.
\end{eqnarray}
Note that the CD follows a (partial) pivotal procedure\cite{Koch03} in which in each iteration a new CD vector (with index $P$) is assigned to the largest of all updated diagonal elements of the two-electron integral matrix with indices $\mu$ and $\nu$. 
The decomposition is continued until the largest updated diagonal element is smaller than a predefined 
Cholesky threshold $10^{-\delta}$. This threshold also determines the accuracy of the decomposition, as it can be shown via the Cauchy-Schwarz inequality that the error of the two-electron integrals, approximated via Eq.~(\ref{cd1}), is in absolute terms always smaller than $10^{-\delta}$.

It has been amply shown\cite{Epifanovsky13,Nottoli21} that the storage requirements for the CVs are substantially lower than for the two-electron integrals so that even for quite large calculations (with more than one thousand basis functions) the whole set of CVs can be kept in core memory unlike the two-electron integrals which either have to be stored on disk or handled using integral-direct algorithms.

For the two-electron integrals in the MO representation, the CVs are transformed from the AO into the MO representation
\begin{equation}
L_{pq}^P = \sum_{\sigma \rho} c_{\sigma p}^* L_{\sigma \rho}^P c_{\rho q}
\end{equation}
such that the MO two-electron integrals are given by
\begin{eqnarray}
(pq|rs) \approx \sum_{P=1}^M L_{pq}^P {L_{sr}^P}^*.
\end{eqnarray}
As for the AO CVs, it is also for the MO CVs usually possible to keep all of them in memory.

Note also that due to the eightfold permutational symmetry of the two-electron integrals the CVs are symmetric with respect an interchange of the two AO or MO indices.

\subsection{\label{sec2c}Cholesky decomposition of the magnetic two-electron integral derivatives}

For derivatives of the two-electron integrals, a CD scheme can be derived by differentiating Eqs.~(\ref{cd1}) and (\ref{cd2}) with respect to the corresponding perturbation.\cite{Feng19} In the case of a magnetic field $\bf B$ as perturbation, this yields
\begin{eqnarray}
\label{cd3}
\frac{\partial (\sigma \rho | \nu \mu)}{\partial B_i} \approx \sum_{P=1}^M \left\{ \frac{\partial L_{\sigma \rho}^P}{\partial B_i} L_{\mu \nu}^P - L_{\sigma \rho}^P \frac{\partial L_{\mu \nu}^P}{\partial B_i}\right\}
\end{eqnarray}
and
\begin{eqnarray}
\label{cd4}
\frac{\partial L_{\sigma \rho}^P}{\partial B_i} &=& \widetilde{(\mu \nu | \nu \mu)}^{- \frac{1}{2}}
\left\{\frac{\partial (\sigma \rho | \nu \mu)}{\partial B_i} \right.\nonumber\\ && \left.- \sum_{R=1}^{P-1} 
\left(\frac{\partial L_{\sigma \rho}^R}{\partial B_i} L_{\mu \nu}^R - L_{\sigma \rho}^R \frac{\partial L_{\mu \nu}^R}{\partial B_i}\right)\right\}.
\end{eqnarray}
Note that there are some differences to the corresponding equations given in Ref.~\onlinecite{Feng19} for nuclear coordinates as perturbation. For a magnetic field the derivative two-electron integrals (and so the derivatives of the CVs) are purely imaginary. As a consequence, the derivatives of the updated diagonal element of the two-electron integral matrix vanishes and Eq.~(\ref{cd3}) consists of the difference instead of the sum of two terms. A similar observation holds for the the correction term to the integral derivative in Eq.~(\ref{cd4}).

Furthermore, the perturbed two-electron integrals no longer exhibit the full eightfold permutational symmetry. However, one can split the perturbed two-electron integrals according to\cite{Kollwitz96}
\begin{eqnarray}
\label{cd5}
\frac{\partial (\sigma \rho | \nu \mu)}{\partial B_i} =
(\frac{\partial \sigma \rho}{\partial B_i} | \nu \mu) +
 (\sigma \rho |  \frac{\partial\nu \mu}{\partial B_i})
\end{eqnarray}
and equate the first (second) term on the right hand side of Eq.~(\ref{cd5}) with the first (second) term on the right hand side of Eq.~(\ref{cd3}). As these partial derivatives in Eq.~(\ref{cd5}) exhibit the full permutational symmetry albeit with an additional consideration of a sign change, one can choose the corresponding perturbed CVs antisymmetric with respect to an AO index change. Eq.~(\ref{cd4}) can be then recast in the following form
\begin{eqnarray}
\label{cd6}
\frac{\partial L_{\sigma \rho}^P}{\partial B_i} &=& \widetilde{(\mu \nu | \nu \mu)}^{- \frac{1}{2}}
\Bigg\{ \frac{1}{2} \left[\frac{\partial (\sigma \rho | \nu \mu)}{\partial B_i} 
+ \frac{\partial (\sigma \rho | \mu \nu)}{\partial B_i} \right]
\nonumber\\ && - \sum_{R=1}^{P-1} 
\frac{\partial L_{\sigma \rho}^R}{\partial B_i} L_{\mu \nu}^R \Bigg\}.
\end{eqnarray}

We emphasize that Eq.~(\ref{cd6}) is not an independent CD of the perturbed integrals, as the Cholesky basis, which is completely determined by the unperturbed two-electron integral matrix, is already defined. In other words, with Eq.~(\ref{cd6}) we are building a representation of the perturbed integrals in a given Cholesky basis.

The perturbed MO two-electron integrals comprise not only the contribution due to the perturbed AO two-electron integrals, but
also contributions due to the perturbed MO coefficients (see Eq.~(\ref{eq12}). It is thus advantageous to define the perturbed CVs in the MO representation as
\begin{eqnarray}
\frac{\partial L_{pq}^P}{\partial B_i} = \sum_{\sigma \rho}\left\{
c_{\sigma p} \frac{\partial L_{\sigma \rho}}{\partial B_i} c_{\rho q}
- \frac{\partial c_{\sigma p}}{\partial B_i}  L_{\sigma \rho} c_{\rho q}
+ c_{\sigma p} L_{\sigma \rho} \frac{\partial c_{\rho q}}{\partial B_i}\right\}\nonumber\\ 
\end{eqnarray}
such that the perturbed MO two-electron integrals can be approximated by
\begin{eqnarray}
\frac{\partial (pq | rs)}{\partial B_i} &\approx &
\sum_{P=1}^M \left\{ \frac{\partial L_{pq}^P}{\partial B_i} L_{sr}^P -
L_{pq}^P \frac{\partial L_{sr}^P}{\partial B_i} \right\} \nonumber\\ & \approx &
\sum_{P=1}^M \left\{ \frac{\partial L_{pq}^P}{\partial B_i} L_{rs}^P + 
L_{pq}^P \frac{\partial L_{rs}^P}{\partial B_i} \right\}.
\end{eqnarray}

\subsection{\label{sec2d}GIAO-MP2 theory with Cholesky-decomposed two-electron integrals}

A CD based implementation of GIAO-MP2 can be carried out by replacing in the expressions given in section~\ref{theor1} all the two-electron integrals by their CD equivalents. In the following, we give the corresponding equations by explicitly replacing the (derivative) integrals with their CD only when this allows for an alternative evaluation of the term; otherwise we will just use $(pq|rs)^\mathrm{CD}$ and ${\partial (pq|rs)^\mathrm{CD}}/{\partial B_i}$ to indicate that the term is evaluated with (derivative) two-electron integrals reconstructed from the CD. For the terms in which we explicitly insert the CD, we indicate a possible way for their evaluation by setting appropriate parentheses.

Before giving the detailed equations, we note that we decided to use in the case of the $(ij|ka)$ as well in most cases for the $(ai|bj)$ integrals (and as well also for the corresponding derivative integrals) the reconstructed integrals. These integrals can be computed from the (derivative) CVs directly after the transformation of the CVs from the
AO into the MO representation. From the $(ai|bj)^\mathrm{CD}$ integrals or from the corresponding derivative integrals
${\partial (ai|bj)^\mathrm{CD}}/{\partial B_i}$, it is possible to obtain the (perturbed) MP2 amplitudes
\begin{eqnarray}
^\mathrm{CD}t_{ij}^{ab} = \frac{(ai|bj)^\mathrm{CD}}{\epsilon_i + \epsilon_j - \epsilon_a - \epsilon_b}
\end{eqnarray}
and
\begin{eqnarray}
{{\partial {^\mathrm{CD}t}_{ij}^{ab}}\over {\partial B_i}} & = & \left\{
{{\partial (ai|bj)^\mathrm{CD}} \over {\partial B_i}}
+ \sum_e \left[ {{\partial f_{ae}} \over {\partial B_i}}\ ^\mathrm{CD}t_{ij}^{eb}
+ {{\partial f_{be}} \over {\partial B_i}}\ ^\mathrm{CD}t_{ij}^{ae} \right] \right. \nonumber \\ && \left.
- \sum_m \left[ ^\mathrm{CD}t_{mj}^{ab} {{\partial f_{mi} }\over {\partial B_i}}
+\ ^\mathrm{CD}t_{im}^{ab} {{\partial f_{mj} }\over {\partial B_i}} \right]
\right\} / (\epsilon_i + \epsilon_j - \epsilon_a - \epsilon_b). \nonumber \\ &&
\end{eqnarray}
Using these amplitudes, one can compute the occupied-occupied and virtual-virtual blocks of the (perturbed) MP2 density matrix according to Eqs.~(\ref{eq3}), (\ref{eq4}), (\ref{eq8}), and (\ref{eq9}). 
For the virtual-occupied block of the (perturbed) density matrix, insertion of the CD yields
\begin{eqnarray}
4 \sum_P L_{ia}^P \left\{ \sum_m \sum_e D_{em} L_{em}^P \right\}
- \sum_P \sum_m L_{ma}^P \left\{ \sum_e D_{em} L_{ie}^P \right\} \nonumber\\
- \sum_P \sum_m L_{im}^P \left\{ \sum_e D_{em} L_{ea}^P \right\} +  D_{ai}
(\epsilon_a - \epsilon_i) 
= - 2\; ^\mathrm{CD}X_{ai}, \nonumber\\
\end{eqnarray}
with
\begin{eqnarray}
^\mathrm{CD}X_{ai} & = & 
\sum_P \sum_{m} \sum_{f} L_{fm}^P \left\{ \sum_e L_{ea}^P\ {^\mathrm{CD}\tilde{t}_{im}^{ef}}^* \right\} \nonumber \\&&
- \sum_{mn} \sum_e (im|en)^\mathrm{CD}\ {^\mathrm{CD}\tilde{t}_{mn}^{ae}}^* \nonumber\\ && 
+ \sum_{mn} D_{mn} \left\{(mn|ia)^\mathrm{CD} - \frac{1}{2} (ma|in)^\mathrm{CD}\right\} \nonumber \\ && 
 + \sum_P L_{ia}^P \left\{ \sum_{ef} D_{ef} L_{ef}^P\right\} - \frac{1}{2} \sum_P \sum_e L_{ea}^p \left\{ \sum_f L_{if}^P D_{ef} \right\}\nonumber\\
 \end{eqnarray}
and
\begin{eqnarray}
&&
 \sum_P \sum_m L_{ma}^P \left\{ \sum_e {{\partial D_{em}} \over { \partial B_i}} L_{ie}^P \right\} 
- \sum_P \sum_m L_{im}^P \left\{ \sum_e {{\partial D_{em}} \over { \partial B_i}} L_{ea}^P \right\} \nonumber\\ && + {{\partial D_{ai}} \over { \partial B_i}}
(\epsilon_a - \epsilon_i) ]
\nonumber \\ & \quad =& - 2 {{\partial ^\mathrm{CD}X_{ai}}\over{\partial B_i}}
+2 \sum_P \frac{\partial L_{ia}^P}{\partial B_i} \left\{ \sum_m \sum_e D_{em} L_{me}^P \right\}\nonumber\\&&
+2 \sum_P \frac{\partial L_{ia}^P}{\partial B_i} \left\{ \sum_m \sum_e D_{em} L_{em}^P \right\}
- \sum_P \sum_m \frac{\partial L_{im}^P}{\partial B_i} \left\{\sum_e L_{ea}^P D_{em} \right\}\nonumber\\&&
- \sum_P \sum_m L_{im}^P \left\{\sum_e \frac{\partial L_{ea}^P}{\partial B_i} D_{em} \right\}
- \sum_P \sum_m \frac{\partial L_{ma}^P}{\partial B_i} \left\{\sum_e L_{ie}^P D_{em}\right\}\nonumber\\&&
- \sum_P \sum_m L_{ma}^P \left\{\sum_e \frac{\partial L_{ie}^P}{\partial B_i} D_{em}\right\}
- \sum_m \sum_e D_{em}\left\{\delta_{im} {{\partial f_{ea}}\over{\partial B_i}} - \delta_{ea}
{{\partial f_{im}}\over {\partial B_i}}\right\} \nonumber\\&&
\end{eqnarray}
with
\begin{eqnarray}
\frac{\partial ^\mathrm{CD}X_{ai}}{\partial B_i} & = & 
\sum_P \sum_{m} \sum_{f} \frac{\partial L_{fm}^P}{\partial B_i} \left\{ \sum_e L_{ea}^P\ {^\mathrm{CD}\tilde{t}_{im}^{ef}}^* \right\} \nonumber \\&&
+ \sum_P \sum_{m} \sum_{f} L_{fm}^P \left\{ \sum_e \frac{\partial L_{ea}^P}{\partial B_i}\ {^\mathrm{CD}\tilde{t}_{im}^{ef}}^* \right\} \nonumber \\&&
+ \sum_P \sum_{m} \sum_{f} L_{fm}^P \left\{ \sum_e  L_{ea}^P\ \frac{\partial {^\mathrm{CD}\tilde{t}_{im}^{ef}}^*}{\partial B_i} \right\} \nonumber \\&&- \sum_{mn} \sum_e \frac{\partial (im|en)^\mathrm{CD}}{\partial B_i}\ {^\mathrm{CD}\tilde{t}_{mn}^{ae}}^* \nonumber\\ && 
- \sum_{mn} \sum_e (im|en)^\mathrm{CD}\ \frac{\partial {^\mathrm{CD}\tilde{t}_{mn}^{ae}}^*}{\partial B_i} \nonumber\\ && 
+ \sum_{mn} \frac{\partial D_{mn}}{\partial B_i} \left\{(mn|ia)^\mathrm{CD} - \frac{1}{2} (ma|in)^\mathrm{CD}\right\} \nonumber \\ && 
+ \sum_{mn} D_{mn} \left\{\frac{\partial (mn|ia)^\mathrm{CD}}{\partial B_i} - \frac{1}{2} \frac{\partial (ma|in)^\mathrm{CD}}{\partial B_i}\right\} \nonumber \\ && 
+ \sum_P L_{ia}^P\left\{ \sum_{ef} \frac{\partial D_{ef}}{\partial B_i}L_{ef}^P\right\}  - \frac{1}{2} \sum_P \sum_e L_{ea}^p \left\{ \sum_f L_{if}^P  \frac{\partial D_{ef}}{\partial B_i} \right\}\nonumber\\&&
 + \sum_P \frac{\partial L_{ia}^P}{\partial B_i} \left\{ \sum_{ef} D_{ef} L_{ef}^P\right\} - \frac{1}{2} \sum_P \sum_e \frac{\partial L_{ea}^p}{\partial B_i} \left\{ \sum_f L_{if}^P D_{ef} \right\}\nonumber\\&&
 + \sum_P L_{ia}^P \left\{ \sum_{ef} D_{ef} \frac{\partial L_{ef}^P}{\partial B_i}\right\} - \frac{1}{2} \sum_P \sum_e L_{ea}^p \left\{ \sum_f \frac{\partial L_{if}^P}{\partial B_i} D_{ef} \right\}·\nonumber\\
\end{eqnarray}

\section{Implementation}

The outlined CD-based GIAO-MP2 approach has been implemented within the CFOUR program package.\cite{cfour,Matthews20c} The
implementation has been carried out with computational efficiency in mind. Thus, we decided to adopt the following guidelines:
\begin{itemize}
    \item the CVs are kept in memory; to be more specific we keep the whole set of unperturbed CVs and one set of perturbed CVs (i.e., the set for one perturbation) simultaneously in memory. Only when the CD of the unperturbed and perturbed two-electron integrals is carried out, we assume that the CVs for all three perturbations can be kept in memory at the same time.
    \item Furthermore, both in memory and on disk we keep at most two vectors of length $n_\mathrm{occ}^2 N_\mathrm{virt}^2$ where 
    $n_\mathrm{occ}$ is the numbers of occupied and $N_\mathrm{virt}$ is the number of virtual orbitals. To be more specific,
    this means that we keep the amplitudes $t_{ij}^{ab}$ and one set of perturbed amplitudes ${\partial t_{ij}^{ab}}/{\partial B_i}$ in memory. The integrals $(ai|bj)$ and ${\partial (ai|bj)}/{\partial B_i}$ are also stored, but they are 
    overwritten by the corresponding amplitudes as soon as the latter are formed. 
    \item To ensure computational efficiency, all terms to be computed have been written,  whenever possible, as matrix-matrix products such that they can be handled via calls to the level 3 BLAS matrix-matrix multiplication routine (DGEMM). This also facilitates shared-memory parallelization via Open MP\cite{openmp} by simply using a threaded BLAS library. 
\end{itemize}
These guidelines suggest that, as in previous GIAO-MP2 implementations,\cite{Gauss92,Kollwitz96} the outer loop in the part where the perturbed density matrices are constructed runs over the components of the external magnetic field. The CD
of the unperturbed and perturbed two-electron integrals have been implemented within the Mainz INTegral (MINT) package\cite{mint} of the CFOUR program package which uses the McMurchie-Davidson scheme\cite{McMurchie78} for
computing integrals. Our implementation of the CD follows more or less the prescriptions given in Ref.~\onlinecite{Koch03} and \onlinecite{Epifanovsky13}. 

Our CD-GIAO-MP2 algorithm is sketched in the flowchart given in Figure~\ref{figure1}. Note that after the transformation of the (perturbed) CVs from the AO into the MO representation the (perturbed) integrals $(ij|ka)^\mathrm{CD}$ and $(ai|bj)^\mathrm{CD}$ are explicitly formed. All other terms are computed as indicated in section~\ref{sec2d}. Note also that both the
HF and CPHF equations are solved using unperturbed and perturbed CVs.

At present our code does not make use of molecular point-group symmetry.

\begingroup
\begin{figure*}[bt] 
\includegraphics[scale=0.60]{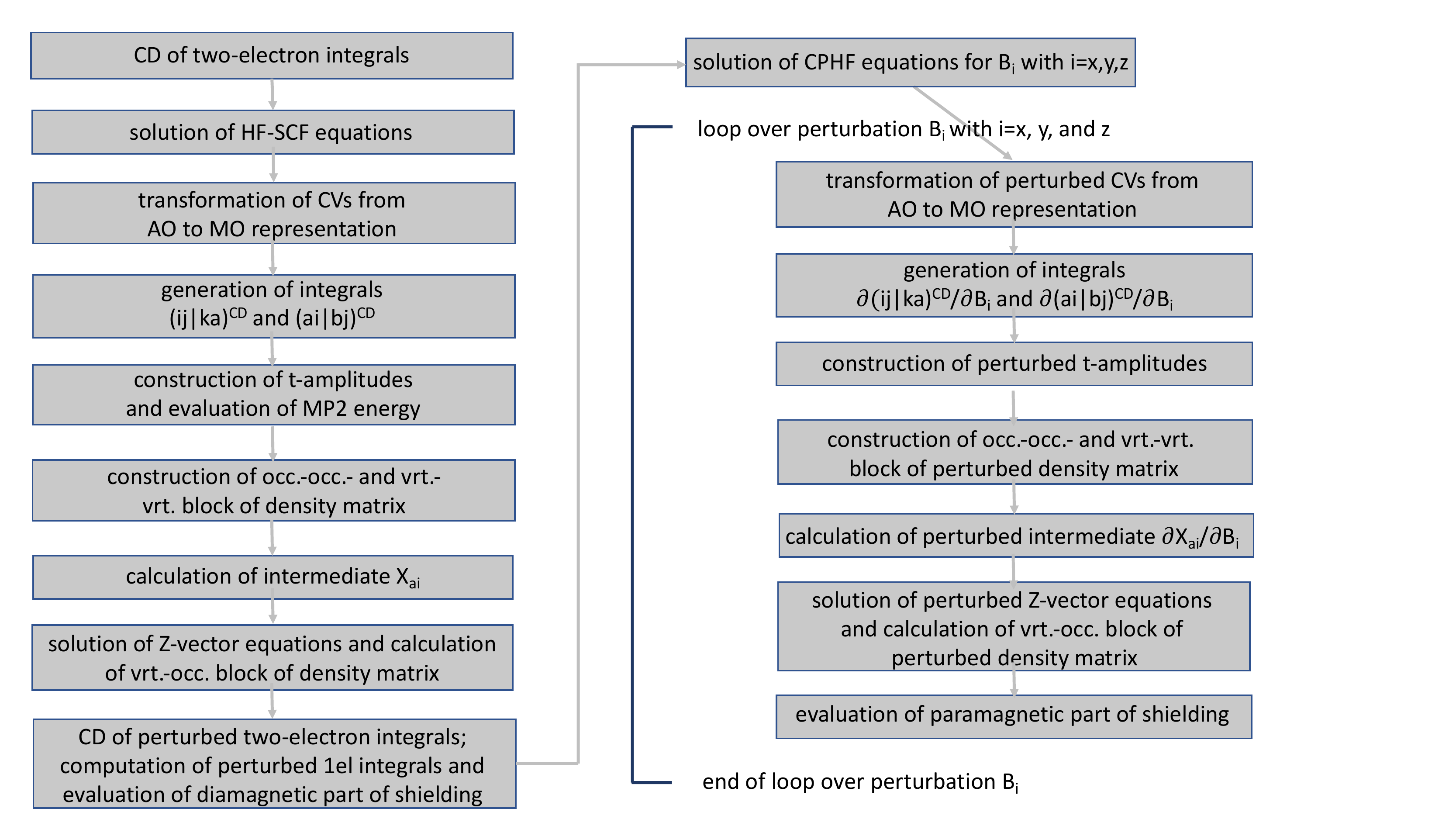}
\caption{Flowchart of a CD based GIAO-MP2 computation.}
\label{figure1}
\end{figure*}
\endgroup

\section{Results}
In the following we will demonstrate the accuracy as well as the applicability of our CD based GIAO-MP2 code. We start 
by investigating the accuracy of the CD for the two-electron integral derivatives (section~\ref{app1}), compare then the accuracy of NMR shieldings computed with CD-GIAO-MP2 with those obtained at the standard GIAO-MP2 level (section~\ref{app2}), and conclude with a few representative examples that are intended to illustrate the applicability of CD-GIAO-MP2 in large-scale computations that otherwise are only feasible in rather costly integral-direct GIAO-MP2 treatments (section~\ref{app3}).

\subsection{Accuracy of the Cholesky decomposition of the derivatives of the two-electron integrals with respect to the components of an external magnetic field}
\label{app1}

In Table~\ref{table1}, we compare the accuracy of the CD of the unperturbed and perturbed two-electron integrals for different Cholesky thresholds $10^{-\delta}$. The computations have been carried out for water (H$_2$O, r(OH)=1.0 \AA, $\langle$(HOH)= 100.0$^{\circ}$) and hydrogen peroxide (H$_2$O$_2$, r(OO) = 1.5 \AA ,
r(OH)= 1.0 \AA, $\langle$(OOH)= 100.0$^{\circ}$, $\tau$(HOOH) = 120.0$^{\circ}$) using Dunning's cc-pVXZ basis sets\cite{Dunning89} with X = D, T, and Q. While for the unperturbed integrals the Cholesky threshold provides a rigorous upper bound for the error in the CD, this is not the case for the perturbed integrals. It is seen that the errors in the perturbed integrals are somewhat larger (by a factor of about 10 to 200) than for the unperturbed integrals and that the error is somewhat larger for calculations with larger basis sets. The 
error due to the CD is furthermore found to be of similar magnitude in the case of H$_2$O and H$_2$O$_2$, thus indicating that the Cholesky threshold will be a useful for judging the accuracy of the unperturbed and perturbed integrals also for larger molecules. 

Overall, we conclude that the Cholesky threshold $\delta$ is a useful measure for the accuracy of the CD in case of the perturbed integrals. \begingroup
\begin{table*}
    \caption{
    Maximum errors (in a.u.) in the computed unperturbed and perturbed two-electron integrals for different CD thresholds $10^{-\delta}$ in calculations for water (H$_2$O) and hydrogen peroxide (H$_2$O$_2$) using the cc-pVXZ basis sets with X = D, T, and Q.}
    \label{table1}
    \centering
    \begin{tabularx}{\textwidth}{p{0.09\textwidth} ccccccccccccccc}
        \hline
         basis set     & two-electron&\multicolumn{12}{c}{Cholesky threshold $\delta$}\\
        & integrals &\quad& 4 &\quad& 5 &\quad& 6 &\quad& 7&&\quad 8 &\quad& 9\\
        \hline
        \multicolumn{4}{l}{a) H$_2$O}\\
        \multirow{2}{*}{cc-pVDZ}	
                & unperturbed && 8.1$\cdot$10$^{-5}$ && 9.5$\cdot$10$^{-6}$ && 1.0$\cdot$10$^{-6}$ && 9.4$\cdot$10$^{-8}$ && 6.4$\cdot$10$^{-9}$ && 8.7$\cdot$10$^{-10}$ \\
            	& perturbed &&3.4$\cdot$10$^{-4}$ && 1.0$\cdot$10$^{-4}$&& 8.6$\cdot$10$^{-6}$&& 2.3$\cdot$10$^{-6}$ && 8.1$\cdot$10$^{-7}$ && 7.4$\cdot$10$^{-8}$ \\
         \multirow{2}{*}{cc-pVTZ}	& unperturbed && 9.5$\cdot$10$^{-5}$ && 9.8$\cdot$10$^{-6}$ && 8.9$\cdot$10$^{-7}$ &&
         8.4$\cdot$10$^{-8}$ &&  8.7$\cdot$10$^{-9}$ &&  1.0$\cdot$10$^{-9}$ \\
            	& perturbed && 3.4$\cdot$10$^{-4}$ && 6.3$\cdot$10$^{-5}$&& 8.4$\cdot$10$^{-6}$&& 1.9$\cdot$10$^{-6}$&&3.1$\cdot$10$^{-7}$&& 9.1$\cdot$10$^{-8}$ \\
         \multirow{2}{*}{cc-pVQZ}	& unperturbed && 9.6$\cdot$10$^{-5}$ && 9.5$\cdot$10$^{-6}$ && 9.6$\cdot$10$^{-7}$ && 9.9$\cdot$10$^{-8}$ && 9.9$\cdot$10$^{-9}$ && 9.6$\cdot$10$^{-10}$ \\
            	& perturbed &&4.9$\cdot$10$^{-4}$ &&4.7$\cdot$10$^{-5}$ && 1.7$\cdot$10$^{-5}$ && 3.3$\cdot$10$^{-6}$&& 9.2$\cdot$10$^{-7}$&& 2.0$\cdot$10$^{-7}$ \\
        \multicolumn{4}{l}{b) H$_2$O$_2$}\\
        \multirow{2}{*}{cc-pVDZ}	
                & unperturbed && 9.4$\cdot$10$^{-5}$ && 9.5$\cdot$10$^{-6}$ && 9.3$\cdot$10$^{-7}$ && 7.9$\cdot$10$^{-8}$ && 8.0$\cdot$10$^{-9}$ && 8.8$\cdot$10$^{-10}$ \\
            	& perturbed &&5.6$\cdot$10$^{-4}$ && 4.4$\cdot$10$^{-4}$&& 6.8$\cdot$10$^{-6}$&& 8.9$\cdot$10$^{-7}$ && 2.9$\cdot$10$^{-7}$ && 5.9$\cdot$10$^{-8}$ \\
         \multirow{2}{*}{cc-pVTZ}	& unperturbed && 9.4$\cdot$10$^{-5}$ && 9.8$\cdot$10$^{-6}$ && 9.9$\cdot$10$^{-7}$ &&
         9.9$\cdot$10$^{-8}$ &&  9.4$\cdot$10$^{-9}$ &&  9.8$\cdot$10$^{-10}$ \\
            	& perturbed && 4.9$\cdot$10$^{-4}$ && 4.5$\cdot$10$^{-5}$&& 5.6$\cdot$10$^{-6}$&& 1.9$\cdot$10$^{-6}$&&4.4$\cdot$10$^{-7}$&& 9.5$\cdot$10$^{-8}$ \\
         \multirow{2}{*}{cc-pVQZ}	& unperturbed && 9.5$\cdot$10$^{-5}$ && 9.9$\cdot$10$^{-6}$ && 9.4$\cdot$10$^{-7}$ && 9.5$\cdot$10$^{-8}$ && 9.9$\cdot$10$^{-9}$ && 9.9$\cdot$10$^{-10}$ \\
            	& perturbed &&5.0$\cdot$10$^{-4}$ &&9.7$\cdot$10$^{-5}$ && 2.2$\cdot$10$^{-5}$ && 6.6$\cdot$10$^{-6}$&& 1.1$\cdot$10$^{-6}$&& 2.3$\cdot$10$^{-7}$\\
        \hline
    \end{tabularx}
\end{table*}
\endgroup
  
\subsection{Accuracy of NMR shieldings computed at the GIAO-MP2 level with Cholesky-decomposed two-electron integrals}
\label{app2}
For the three organic molecules acetaldehyde (H$_3$C-CHO), vinyl alcohol (H$_2$C=CHOH), and ethylene oxide (C$_2$H$_4$O) we compare in Table~\ref{table2} the results from CD-based GIAO-HF and GIAO-MP2 computations with those from standard GIAO-HF and GIAO-MP2 treatments, again for different Cholesky thresholds $\delta$ and for different basis sets from Dunning's cc-pVXZ hierarchy with X = D, T, and Q. The geometries of these molecules have been determined at the same level as the NMR shieldings and are given for completeness in the \hyperlink{si}{supplementary material}. Table~\ref{table2} reports the corresponding maximum absolute errors in the isotropic shieldings obtained in the CD based computations. One can conclude from the data in Table~\ref{table2} that the corresponding errors are for all Cholesky thresholds small and of no relevance for actual computations. The error amounts to several hundredths ppm in the case of $\delta$=4, to about a few thousandths ppm in the case of $\delta$=5, and to less than a thousandth ppm in the case of $\delta=6$.
\begingroup
\begin{table*}
    \caption{
    Maximum absolute errors (given in ppm) in the computed isotropic shieldings (with respect to a standard GIAO-HF and GIAO-MP2 computation) for acetaldehyde, ethylene oxide, and vinyl alcohol obtained in CD based GIAO-HF and GIAO-MP2 computations for different Cholesky thresholds $\delta$ using the cc-pVXZ basis sets with X = D, T, and Q.}
    \label{table2}
    \centering
    \begin{tabularx}{\textwidth}{p{0.12\textwidth} lcccccccccccc}
        \hline
        nucleus  && \multicolumn{3}{c}{$\delta$=4} &\qquad\qquad&\multicolumn{
        3}{c}{$\delta$=5} &\\ \cline{3-5} \cline{7-9}
        && cc-pVDZ & cc-pVTZ & cc-pVQZ  &\qquad& cc-pVDZ & cc-pVTZ & cc-pVQZ \\
        \hline
        \multicolumn{9}{l}{a) HF treatment}\\
       	
                  $^{13}$C &&0.013 &0.002& 0.008  && 0.002 & 0.001 & 0.001 \\
                  $^{17}$O && 0.063 & 0.007 & 0.021 && 0.001 & 0.004 & 0.001\\
            	  $^{1}$H && 0.001 & 0.001 & 0.001 && 0.000 & 0.000 & 0.001 \\
            \multicolumn{9}{l}{b) MP2 treatment} \\
                  $^{13}$C &&0.013 &0.002 & 0.081 && 0.001 & 0.002 & 0.002 \\
            	  $^{17}$O && 0.049 & 0.015 & 0.037 && 0.003 & 0.002 & 0.003 \\
            	  $^{1}$H && 0.001 & 0.001 & 0.001 && 0.001 & 0.001 & 0.000 \\
                 \hline
    \end{tabularx}
\end{table*}
\endgroup
The comparison suggests that a Cholesky threshold of $\delta =5$ is sufficient to guarantee converged values for the shieldings, though the results obtained with 
$\delta$=4 already exhibit an accuracy that is fully acceptable for chemical applications. 

At this point, it is interesting to compare the accuracy of the present CD based scheme with the accuracy of corresponding schemes employing density fitting. 
According to Ref.~\onlinecite{Stoychev18a} the maximum absolute error due to density fitting in GIAO-MP2 computations with the pcSseg3 basis~\cite{Jensen15} amounts in chemical shift computations to about 0.035 ppm (carbon shieldings), 0.055 ppm (nitrogen, oxygen, fluorine, phosphorus shieldings), and 0.002 ppm (hydrogen shieldings) which roughly corresponds to the accuracy obtained with $\delta=4$. However, the advantage of the present CD scheme is that the error can be easily controlled via the choice of the Cholesky threshold $\delta$, while this is not the case when using density fitting. The error is there due to the choice of the auxiliary basis and it is hard to improve the results in a systematic manner.

One might also consider a mixed scheme that combines a rigorous HF treatment (without CD) with a CD based MP2 part. 
In this way, one may hope to reduce the error 
when using a rather loose Cholesky threshold (i.e., $\delta$=4). 
However, based on results from some exploratory calculations, it is concluded that such a scheme offers no significant improvement for the computation of NMR shieldings.

\begingroup
\begin{figure*}[bt] 
\includegraphics[scale=0.60]{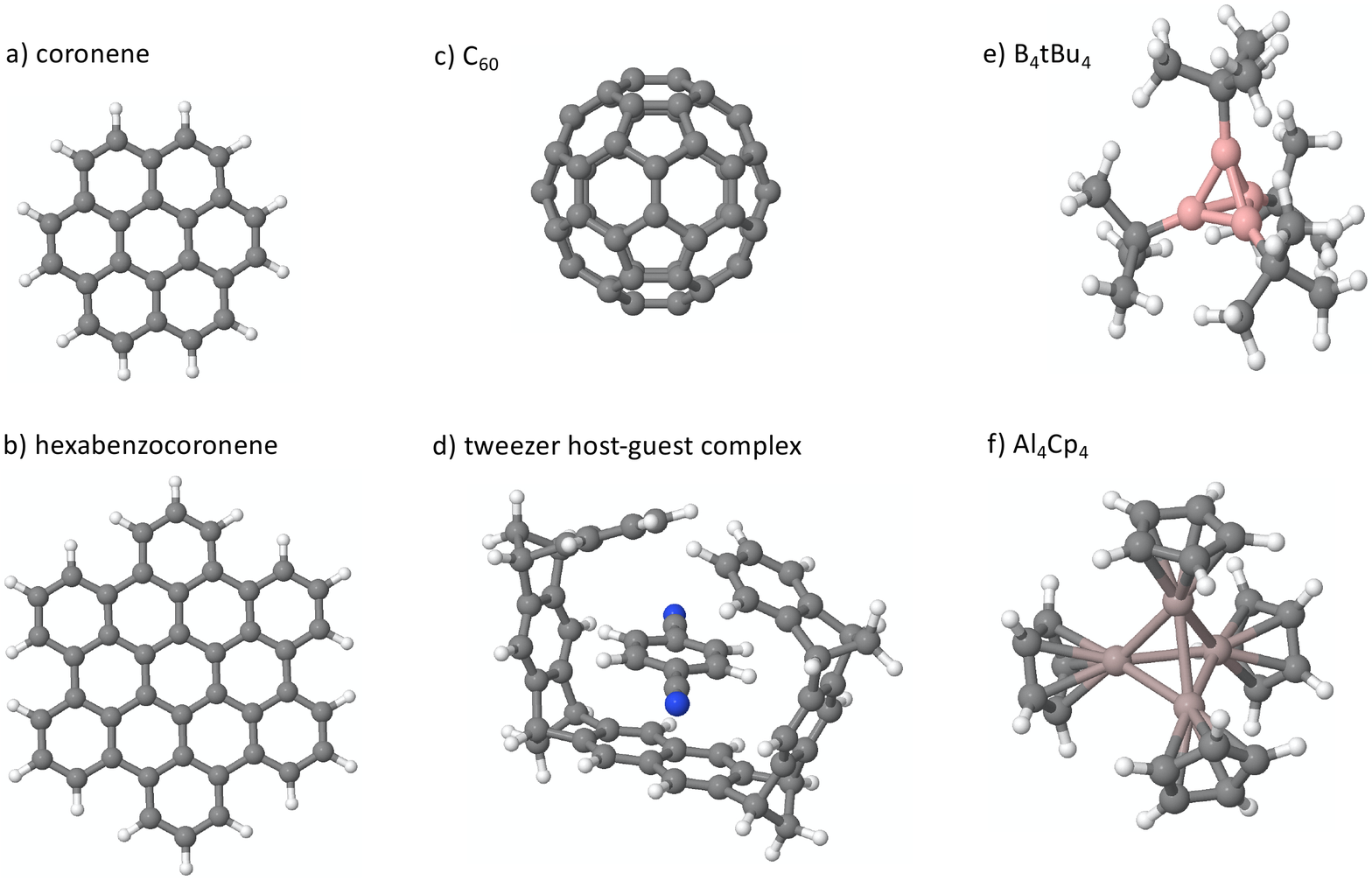}
\caption{Structures of the molecules for which representative CD based GIAO-MP2 computations have been carried out.
In the ball-and-stick representations, grey balls represent carbon, white balls hydrogen, blue balls nitrogen, pink balls boron, and violet balls aluminum.}
\label{figure2}
\end{figure*}
\endgroup

\subsection{Representative Applications}
\label{app3}
To illustrate the applicability of our CD based GIAO-MP2 computations, we report in the following results from corresponding calculations for several large molecules with up to close to 100 atoms and more than 1000 basis functions. 
The chosen examples comprise coronene (C$_{24}$H$_{12}$), hexabenzocoronene (C$_{42}$H$_{18}$), tetrakis(t-butyl)tetraborane(4) (B$_4$C$_{16}$H$_{36}$), tetrameric cyclopentadienyl aluminum(I) (Al$_4$C$_{20}$H$_{20}$), the buckminsterfullerene (C$_{60}$), and a tweezer host-guest complex (C$_{54}$N$_2$H$_{36}$). The structures
of these molecules are depicted in Figure~\ref{figure2} with the corresponding Cartesian coordinates given in the \hyperlink{si}{supplementary material}. The geometries for the calculations have been taken from Ref.~\onlinecite{Stoychev21} in the case of coronene and the tweezer host-guest complex, from Ref.~\onlinecite{Ochsenfeld01} in the case of hexabenzocoronene, from Ref.~\onlinecite{Kollwitz98} in the case of B$_4$$t$-Bu$_4$ as well as Al$_4$Cp$_4$, and from Ref.~\onlinecite{Haeser89} in the case of C$_{60}$.
For a detailed account why NMR chemical-shift computations for these molecules are important, we refer in the case of hexabenzocoronene to Ref.~\onlinecite{Ochsenfeld01}, in the case of the tetrahedral boron compound to Ref.~\onlinecite{Buehl93}, in the case of the tetrameric aluminum(I) compound to Ref.~\onlinecite{Gauss93a}, and for the tweezer host-guest complex to Ref.~\onlinecite{Brown01}. 

The calculations have been performed with the dzp (11s7p1d/6s4p1d for Al, 8s4p1d/4s2p1d for B, C, O as well as N, and 4s1p/2s1p for H) and tz2p (12s9p2d/7s5p2d for Al, 9s5p2d/5s3p2d for B, C, O, as well as N, and 5s2p/3s2p for H) versions of the Karlsruhe basis set\cite{Schaefer92} and polarization functions from Ref.~\onlinecite{Gauss93}; only for the rather large tweezer host-guest complex the larger calculation had to be restricted to tzp (9s5p1d/5s3p1d for C as well as N and 5s1p/3s1p for H) instead of tz2p.
For completeness, the basis sets as well as the results of the calculations, i.e., the obtained shieldings are given in the \hyperlink{si}{supplementary material}, while the focus in the following is on the size of these computations and the computational resources required to perform the computations.

In Table~\ref{table3} we report the number of basis functions ($N_{\mathrm{bf}}$), the number of CVs ($N_{\mathrm{CV}}$), and the CD compression rate (ratio of maximum number and actual number of CVs) for the computations with two different Cholesky thresholds ($\delta=4$ and $\delta=5$).
We also report the timings of the
calculations together with the memory requirements. As seen from the table, the required resources are still rather modest for calculations with a couple of hundred basis functions, while they are quite demanding for the computations with more than 1000 basis functions. This finding indicates that the application range of our CD-GIAO-MP2 approach comprises cases with several hundred up to one thousand basis functions, while calculations with significantly more than 1000 basis function require additional means to exploit sparsity and the local nature of electron correlation and thus are better treated using, for example, the recently reported DLPNO-MP2 approach\cite{Stoychev21} for computing NMR shieldings.

A closer look at Table~\ref{table3} reveals that the CD very effectively reduces the memory requirements for the two-electron integrals. The CD compression rates are in all cases high (i.e., above 28) and reach for the most demanding computations on C$_{60}$ and the tweezer host-guest complex in the case of $\delta$=5 values of 94 and 98, respectively. The actual number of CVs is in all cases merely several thousand with only for C$_{60}$ when using the tz2p basis and a threshold of 5 reaching a value above 10000. For a threshold of 4, the number of CVs is typically 
by about 20 to 25 \% lower and the compression rates are higher by the same percentage.
For C$_{60}$, this means that for $\delta=4$ only 8776 CVs are required instead of the 11022 CVs for $\delta=5$.

For the calculation on C$_{60}$ with the tz2p basis and $\delta=5$, the memory that is required to keep the whole set of CVs in core is about 86 GB for the unperturbed two-electron integrals and about 256 GB for the perturbed two-electron integrals, i.e., amounts of memory that nowadays are easily affordable. In the case of $\delta=4$, these values reduce to about 68 GB for the unperturbed and to about 204 GB for the perturbed integrals. Loosening the Cholesky threshold does not only reduce the memory requirements but also speeds up the calculation (at least those steps whose cost scales with the number of CVs). 
For the large calculation on C$_{60}$ with the tz2p basis, this means that the memory requirement is reduced from 1089 GB ($\delta=5$) to 1023 GB $(\delta=4$) and the calculation only required 3 days, 9 hours, and 49 minutes instead of 4 days, 11 hours, and 31 minutes (see table~\ref{table3}). The computational results differed by at most 0.02 ppm which is negligible for chemical applications.

The limiting factor of GIAO-MP2 computations using the currently implemented CD based scheme is the necessity to store for the MP2 treatment two vectors of length $N_\mathrm{virt}^2 n_\mathrm{occ}^2$ in memory (one vector requires about 384 GB in the case of C$_{60}$ and the tz2p basis). For that reason the largest computations (on the tweezer host-guest complex and C$_{60}$, both with more than 1000 basis functions) needed 814 or even 1089 GB core memory, respectively. However, the actual computational times (we report wall-clock times) are in all cases acceptable with only the largest computations requiring more than a day. The tz2p computation for C$_{60}$ required about 4 days and 12 hours when using 20 cores with $\delta =5$ and  3 days and 10 hours with $\delta =4$, while the tzp calculation for the tweezer host-guest complex has been completed, again using 20 cores, in slightly less than 2 days and 10 hours. We note that parallelization is essential for ensuring reasonable wall-clock timings, even though the parallelization efficiency is unfortunately not very high. For example, the dzp computation on the tweezer host-guest complex required about 5 days when running on one CPU and only 1 day and 5 hours when running on 20 CPUs. The speed up of about 4.1 is not too convincing, but we note that at this stage no specific effort was put into achieving an efficient parallelization, as we only used threaded matrix-matrix multiplication routines from an appropriate BLAS library. 
We are convinced that with additional effort a higher parallelization speed up can be reached, but this is an issue beyond the scope of the present paper.

In Table~\ref{table4}, we present a detailed analysis of the computational cost for the dzp and tzp computations on the tweezer host-guest complex. As it is seen, the costs are dominated by the following
steps: (a) the CD of the (perturbed) two-electron integrals, all SCF related steps (solution of the HF-SCF, CPHF, as well as unperturbed and perturbed Z-vector equations), the construction of unperturbed and perturbed $(ai|bj)$ integrals, and the unperturbed and perturbed $(ab|ci)$ contributions to the unperturbed and perturbed $X_{ai}$ intermediate. These findings are expected, as the costs of these steps are those that in a formal analysis appear the most costly. The cost for the CD scales as $N_{\mathrm{CV}}^2N_{\mathrm{bf}}^2$, the SCF related steps exhibit a scaling of $N_{\mathrm{bf}}^2 n_{\mathrm{occ}}N_\mathrm{CV}$ (SCF) or $N_{\mathrm{virt}}^2 n_{\mathrm{occ}}N_\mathrm{CV}$ (CPHF, Z-vector equations), respectively, due to the exchange contributions, the formation of the $(ai|bj)$ integrals scales as $N_\mathrm{CV} N_\mathrm{virt}^2n_\mathrm{occ}^2$, and the cost for the $(ab|ci)$ contributions to the $X_{ai}$ intermediates are of the same order. The efficiency of a CD based calculation thus depends in a rather crucial manner on the number of Cholesky vectors used for the representation of the unperturbed and perturbed two-electron integrals. 

\begingroup
\begin{table*}
\caption{Computational requirements for the CD based GIAO-MP2 computations (with Cholesky thresholds $\delta$=4 and $\delta$=5) for the molecules given in Fig.~\ref{figure2}. The number of basis functions is denoted by $N_\mathrm{bf}$, $N_\mathrm{el}$ is the number of electrons, $N_{CPU}$ specifies the number of CPUs used, and $N_\mathrm{CV}$ the number of Cholesky vectors in the corresponding calculation. The compression is defined as the ratio of the theoretically maximum number of Cholesky vectors and the actual number $N_\mathrm{CV}$. The required memory is given in GB and the wall-clock time $t_{wall}$ in terms of days, hours, and minutes.
    If not otherwise noted, calculations have been carried out on an Intel Xeon(R) E5-2643 node running at 3.4 GHz}
    \label{table3}
    \centering
    \begin{tabularx}{\textwidth}{p{0.150\textwidth} cccccccccccccccccccccccc}
        \hline
        \multirow{3}{*}{molecule} & \multirow{3}{*}{basis} &\quad& \multirow{3}{*}{$N_\mathrm{bf}$}&\quad& \multirow{3}{*}{$N_\mathrm{el}$} &\quad& \multirow{3}{*}{$N_\mathrm{CPU}$} &  &\multicolumn{7}{c}{$\delta$=4} &&\multicolumn{7}{c}{$\delta$=5} \\ \cline{10-16} \cline{18-24} &&&&&&&&\quad& $N_\mathrm{CV}$  &\quad& compression & \quad& memory &\quad& $t_\mathrm{wall}$ &\quad& $N_\mathrm{CV}$  &\quad& compression & \quad& memory &\quad& $t_\mathrm{wall}$\\
                       &&&&&&\quad& &\quad&  &\quad& &\quad&  [GB]  &\quad& [d:h:min] &\quad& &\quad&&\quad&  [GB]  &\quad& [d:h:min]  \\
       \hline
        \multirow{2}{*}{coronene}	
& dzp  &&420 && 156 && 8&& 2493 &&35.46  && 17 && 1:05 &&3097&&28.55 && 19 && 1:20\\
& tz2p  && 684 && 156&& 8 && 3917 && 59.81 && 60 && 3:50 &&4959&&47.24&& 67 && 4:50\\
          \multirow{2}{*}{hexabenzocoronene} &dzp  &&720 && 270 && 8 && 4310 && 60.22 && 124 && 5:38 &&5331 &&48.69 && 132  && 8:28\\ 
                            &tz2p &&1170 && 270 && 8 && 6710 && 102.09 && 422 && 1:01:28 &&8546 &&80.16 &&457 && 1:07:35 \\
          \multirow{2}{*}{B$_4$tBu$_4$}             &dzp  &&480 &&152 && 8 && 2627 && 43.94 && 23 && 1:15 &&3402 &&33.93 && 25 && 1:22\\
                            &tz2p&& 804 &&152 && 8 && 4281&& 75.59&& 86 && 4:51 && 5511 &&58.72 && 97 && 5:57 \\
          \multirow{2}{*}{Al$_4$Cp$_4$}            &dzp  &&492 && 192  &&8 &&2738&& 44.29 && 31 && 1:35 &&3494 &&34.71 && 34 && 1:58\\
                            &tz2p &&788 && 192 && 8&& 4414 && 70.43 && 105 && 5:37&& 5524 &&56.28 && 115 && 6:56 \\
          \multirow{2}{*}{C$_{60}$ }              &dzp  &&900  &&360 && 8 &&5502 &&73.69 && 312 && 19:07 &&6722 &&60.32 && 325 &&  23:20\\
                            &tz2p &&1440 &&360 && 20 && 8776 && 118.22 && 1023 &&3:09:49$^a$&& 11022 &&94.13 && 1089 && 4:11:31$^a$\\
          tweezer host-          &dzp  &&1020 && 374  && 20 && && && && &&7400 &&70.37 &&  649 && 1:06:11$^a$\\
            guest complex                 &tzp  &&1280 && 374 && 20 && && &&  && &&8344 &&98.26 && 814 &&  2:09:26$^a$ \\
        \hline
    \end{tabularx}
    $^a$ calculation has been carried out on an Intel Xeon(R) Gold 5215M node running at 2.5 GHz.\\
\end{table*}
\endgroup
\begingroup
\begin{table*}
    \caption{Breakup of the computational timings (in minutes) for the CD based GIAO-MP2 computations (with $\delta$=5) for the tweezer host-guest complex using a dzp and tzp basis, respectively.
    Calculations have been carried out on an Intel Xeon(R) Gold 5215M node running at 2.5 GHz}
    \label{table4}
    \begin{center}
    \begin{tabular}{lcccccccccccccccc}
        \hline
        computational  step &\quad&  dzp basis$^a$ &\quad& dzp basis$^b$&\quad& tzp basis$^b$\\         
       \hline
       CD of two-electron integrals &&  163 &&  33 && 60 \\
       solution of HF-SCF equations && 221 && 57 && 208 \\
       transformation of unperturbed CVs into MO basis && 33 && 7  && 19 \\
       construction of integrals $(ij|ka)$ && 34 && 4 && 6 \\
       construction of integrals $(ai|bj)$ && 261 && 35 && 67 \\
       construction of t amplitudes && 11 && 1  && 1 \\
       computation of occ.-occ. block of density matrix && 11 && 1 && 2 \\
       computation of vrt.-vrt. block of density matrix && 51 && 5 && 12\\
       evaluation of the contribution due to $(ij|ka)$ to $X_{ai}$ &&  35&& 4 && 8  \\
       evaluation of the contribution due to $(ab|ci)$ to $X_{ai}$ && 268 && 51 && 99 \\
       solution of Z-vector equation && 52  && 17 && 37 \\
       CD of perturbed two-electron integrals && 805 && 125 && 230 \\
       solution of CPHF equations && 101 && 24 && 46\\
       transformation of perturbed CVs with unperturbed MOs && 3 x 33 && 3 x 7  && 3 x 21 \\
       transformation of unperturbed CVs with perturbed MOs && 3 x 64 && 3 x 12  && 3 x 32 \\
       construction of integrals $\partial(ij|ka)/\partial B_i$ && 3 x 91 && 3 x 12 && 3 x 18 \\
       construction of integrals $\partial(ai|bj)/\partial B_i$ && 3 x 514 && 3 x 69 && 3 x 134 \\
       evaluation of the contribution due to $\partial (ij|ka)/\partial B_i$ to $\partial X_{ai}/\partial B_i$ && 3 x 25 && 3 x 4 &&3 x 8.3 \\
       evaluation of the contribution due to $\partial(ab|ci)/\partial B_i$ to $\partial X_{ai}/\partial B_i$ && 3 x 523 && 3 x 98 &&3 x 191 \\
       construction of perturbed t amplitudes && 3 x 68 && 3 x 7 && 3 x 15 \\
       computation of occ.-occ.block of perturbed density matrix && 3 x 12 && 3 x 1 && 3 x 3 \\
       computation of vrt.-vrt. block of perturbed density matrix && 3 x 55 && 3 x 10  && 3 x 25 \\
       evaluation of the contribution due to $(ij|ka)$ to $\partial X_{ai}/\partial B_i$ && 3 x 58 && 3 x 4  && 3 x 8\\
       evaluation of the contribution due to $(ab|ci)$ to $\partial X_{ai}/\partial B_i$ && 3 x 264 && 3 x 49  &&3 x 97  \\       
        solution of perturbed Z-vector equations && 3 x 27 && 3 x 2 &&3 x 17  \\
       total wall-clock time && 7215 && 1740 && 3446 \\
         \hline \\
    \end{tabular}
           \end{center}
     $^a$ calculation has been performed using 1 CPU \\
    $^b$ calculation has been performed using 20 CPUs; CPU time is given per node \\
    \end{table*}
\endgroup

Finally, we note that in case of coronene and the dzp basis (420 basis functions) the CD based GIAO-MP2 scheme outperforms the standard GIAO-MP2 approach based on the regular two-electron integrals. The CD based calculation (with $\delta$=4) runs in 2 hours and 5 minutes, while the standard GIAO-MP2 calculations requires 9 hours and 42 minutes (both calculations have been carried out on 1 CPU of an Intel Xeon(R) Gold 5215M node running at 2.5 GHz). This clearly shows that CD based GIAO-MP2 computations with several hundred basis functions are already beyond the break-even point at which the CD based schemes surpasses the standard approach in terms of computational efficiency.

\section{Conclusions and outlook}

In this paper, we report on an MP2 scheme for the computation of NMR shieldings that uses a Cholesky decomposition for the handling of the unperturbed and perturbed two-electron integrals. The latter arise when using GIAOs for ensuring gauge-origin independence. The storage and handling of these integrals is the
main bottleneck\cite{Kollwitz96} of traditional electron-correlated NMR chemical-shift computations and has significantly limited their applicability in the past. As shown, the CD for the derivatives of the two-electron integrals with
respect to the components of an external magnetic field can be achieved by using a recipe obtained
by straightforward differentiation of the relevant equations for the Cholesky decomposition of the corresponding unperturbed
integrals. Such a scheme has already been successfully applied for geometrical derivatives\cite{Feng19} of 
the two-electron integrals and is used here for the first time for an external magnetic field as perturbation. As shown, the CD of the perturbed two-electron integrals leads to a very compact representation that allows to maintain the whole set of unperturbed and perturbed Cholesky vectors 
in core memory, even for very large calculations.
We have derived the required formulae for a CD-based GIAO-MP2 approach and have reported on its first implementation. Calculations on systems consisting of close to one hundred atoms and with more than one thousand basis functions demonstrate the applicability of our scheme. 
However, for even larger systems it becomes mandatory to couple the present CD based scheme with techniques that enable a local treatment of correlation\cite{Gauss00a,Loibl12,Stoychev21} and that allow to exploit sparsity. 

The present paper represents a first step along the lines of formulating and implementing electron-correlated approaches for the computation of magnetic properties for large molecules. Further work can be envisioned and is planned in two directions.
The first is to move to a multiscale description for computing NMR chemical shifts\cite{Karadakov00,Sebastiani04,Gascon05,Lipparini13,Chung15,Jin18,Hashem20} in which the present CD based 
GIAO-MP2 scheme provides the engine for the treatment of the QM region. 
MP2 fails in the accurate prediction of absolute shieldings (it has a strong tendency to overshoot\cite{Gauss96a}) and its performance deteriorates in cases where electron-correlation effects are large (see, for example, Ref.~\onlinecite{Stanton96a}). Therefore, the second direction is an extension
of the present scheme towards coupled-cluster methods,\cite{Gauss95a,Gauss95b,Gauss96a} as it has been amply shown that highly accurate results are only achieved using these methods.\cite{Gauss02c}

\section{DATA AVAILABLITY STATEMENT}

The data that supports the findings of this study are available within the article and its \hyperlink{si}{supplementary material}.

\section{Supplementary material}

See \hyperlink{si}{supplementary material} for the details of the reported calculations (geometries in Cartesian Coordinates and basis sets) as well the CD based GIAO-MP2 results.

\begin{acknowledgments}
One of the authors (J.G.) thanks Professor Anna I. Krylov (University of Southern California, Los Angeles, USA) for introducing him to the Cholesky decomposition as well as for encouragement to implement Cholesky decomposition into the CFOUR program package and to pursue research along these lines. S.S. acknowledges support from the Deutsche Forschungsgemeinschaft via grant STO 1239/1-1.
\end{acknowledgments}
\label{sec:cc}

\bibliography{cdgiaomp2}

\end{document}



\title{Supporting Information: NMR Chemical Shift Computations at Second-Order Møller-Plesset Perturbation Theory
Using Gauge-Including Atomic Orbitals and Cholesky-Decomposed Two-Electron Integrals}

\author{Sophia Burger}
 \email{sburger@students.uni-mainz.de}
 \affiliation{Department Chemie, Johannes Gutenberg-Universit{\"a}t Mainz, Duesbergweg 10-14, D-55128 Mainz, Germany}
\author{Filippo Lipparini}%
 \email{filippo.lipparini@unipi.it}
\affiliation{Dipartimento di Chimica e Chimica Industriale, Universit\`{a} di Pisa, Via G. Moruzzi 13, I-56124 Pisa, Italy}
\author{J{\"u}rgen Gauss}%
 \email{gauss@uni-mainz.de}
\affiliation{Department Chemie, Johannes Gutenberg-Universit{\"a}t Mainz, Duesbergweg 10-14, D-55128 Mainz, Germany}
\author{Stella Stopkowicz}%
 \email{stella.stopkowicz@uni-mainz.de}
\affiliation{Department Chemie, Johannes Gutenberg-Universit{\"a}t Mainz, Duesbergweg 10-14, D-55128 Mainz, Germany}
\
%
\date{\today}

%
%
%
%
%
%
%

\date{\today}

\maketitle

\section{Computational details and results of the GIAO-MP2 computations for the Validation of the CD based GIAO-MP2 scheme}

In this section, we provide the geometries (Cartesian coordinates in bohr) of the three organic molecules acetaldehyde, ethylene oxide, and vinyl alcohol that has been used in the computations to validate the accuracy of the CD based GIAO-MP2 scheme
together with the computed shieldings at the standard GIAO-MP2 level and the CD based GIAO-MP2 level for different Cholesky thresholds.

\subsection{geometries}

\noindent
a) acetaldehyde
\begin{verbatim}
MP2/cc-pVDZ
     O        -2.17634543     0.43776210     0.00000000
     C        -0.23383524    -0.79604444     0.00000000
     C         2.39046218     0.31579670     0.00000000
     H        -0.29575634    -2.91253183     0.00000000
     H         2.29268343     2.38934380     0.00000000
     H         3.43233336    -0.35309330     1.67586465
     H         3.43233336    -0.35309330    -1.67586465
\end{verbatim}
\begin{verbatim}
MP2/cc-pVTZ
     O        -2.15713774     0.43383064     0.00000000
     C        -0.22691728    -0.79082691     0.00000000
     C         2.36701049     0.31421146     0.00000000
     H        -0.30433891    -2.86824346     0.00000000
     H         2.27323815     2.35673730     0.00000000
     H         3.39235944    -0.34936120     1.65116086
     H         3.39235944    -0.34936120    -1.65116086
\end{verbatim}
\begin{verbatim}
MP2/cc-pVQZ
     O        -2.15406416     0.43468920     0.00000000
     C        -0.22741238    -0.79104713     0.00000000
     C         2.36452175     0.31401372     0.00000000
     H        -0.31397604    -2.87136370     0.00000000
     H         2.27555986     2.35639406     0.00000000
     H         3.38939116    -0.35195411     1.64994995
     H         3.38939116    -0.35195411    -1.64994995
\end{verbatim}

\noindent
b) ethylene oxide
\begin{verbatim}
MP2/cc-pVDZ
     O         0.00000000     0.00000000     1.51839207
     C         0.00000000    -1.39159763    -0.80499899
     C         0.00000000     1.39159763    -0.80499899
     H        -1.75381137    -2.40501407    -1.23200344
     H         1.75381137    -2.40501407    -1.23200344
     H         1.75381137     2.40501407    -1.23200344
     H        -1.75381137     2.40501407    -1.23200344
\end{verbatim}
\begin{verbatim}
MP2/cc-pVTZ
     O         0.00000000     0.00000000     1.51336764
     C         0.00000000    -1.37716295    -0.80401632
     C         0.00000000     1.37716295    -0.80401632
     H        -1.72980116    -2.36923116    -1.21791839
     H         1.72980116    -2.36923116    -1.21791839
     H         1.72980116     2.36923116    -1.21791839
     H        -1.72980116     2.36923116    -1.21791839
\end{verbatim}
 \begin{verbatim}
MP2/cc-pVQZ
     O         0.00000000     0.00000000     1.51266217
     C         0.00000000    -1.37482389    -0.80454204
     C         0.00000000     1.37482389    -0.80454204
     H        -1.73050490    -2.36989164    -1.21198945
     H         1.73050490    -2.36989164    -1.21198945
     H         1.73050490     2.36989164    -1.21198945
     H        -1.73050490     2.36989164    -1.21198945
\end{verbatim}
 
\noindent
c) vinyl alcohol
\begin{verbatim}
MP2/cc-pVDZ
     O        -2.19512947     0.25637924     0.00000000
     C         0.14503021    -0.81329587     0.00000000
     C         2.38619652     0.39199311     0.00000000
     H        -1.95424727     2.07196077     0.00000000
     H         0.00750482    -2.87558531     0.00000000
     H         4.12975399    -0.70460982     0.00000000
     H         2.51640305     2.45568961     0.00000000
\end{verbatim}
\begin{verbatim}
MP2/cc-pVTZ
     O        -2.17601920     0.25117592     0.00000000
     C         0.15547292    -0.81118208     0.00000000
     C         2.35803138     0.39471773     0.00000000
     H        -1.97151641     2.05813197     0.00000000
     H         0.03076188    -2.83910208     0.00000000
     H         4.08118218    -0.67297068     0.00000000
     H         2.46671106     2.42636625     0.00000000
\end{verbatim}
 \begin{verbatim}
MP2/cc-pVQZ
     O        -2.17242029     0.25144895     0.00000000
     C         0.15607466    -0.81229904     0.00000000
     C         2.35425421     0.39573093     0.00000000
     H        -1.98270474     2.05640609     0.00000000
     H         0.02748792    -2.84295692     0.00000000
     H         4.07830975    -0.67330793     0.00000000
     H         2.46473782     2.42918651     0.00000000
\end{verbatim}

\subsection{isotropic NMR shieldings}

\noindent
a) acetaldehyde
\\

\noindent
Cholesky threshold = 4
\begin{verbatim}
HF/cc-pVDZ
       O             -357.689         
       C                2.778           
       C              171.860            
       H               22.483             
       H               29.599             
       H               29.793             
       H               29.793             
\end{verbatim}
\begin{verbatim}
MP2/cc-pVDZ
       O             -253.816           
       C               31.263           
       C              175.841            
       H               22.224            
       H               29.734             
       H               29.560             
       H               29.560            

\end{verbatim} 
\begin{verbatim}
HF/cc-pVTZ
       O             -360.523          
       C               -8.788           
       C              165.184            
       H               22.617             
       H               30.045             
       H               30.166             
       H               30.166             
\end{verbatim}
\begin{verbatim}
MP2/cc-pVTZ
       O             -284.728           
       C                8.274           
       C              166.744            
       H               22.150             
       H               29.980             
       H               29.780             
       H               29.780             
 \end{verbatim} 
\begin{verbatim}
HF/cc-pVQZ
       O             -360.066          
       C              -13.040           
       C              163.213            
       H               22.510             
       H               29.958             
       H               30.084             
       H               30.084             
\end{verbatim}
\begin{verbatim}
MP2/cc-pVQZ
       O             -282.404       
       C                1.634           
       C              163.972            
       H               21.932             
       H               29.790             
       H               29.608             
       H               29.608             
 \end{verbatim} 

 \noindent
   Cholesky threshold = 5
 \begin{verbatim}
HF/cc-pVDZ       
       O             -357.734         
       C                2.771           
       C              171.846            
       H               22.482             
       H               29.600             
       H               29.792             
       H               29.792             
 \end{verbatim}
 \begin{verbatim}
MP2/cc-pVDZ
       O             -253.780           
       C               31.267           
       C              175.829            
       H               22.224             
       H               29.734             
       H               29.560             
       H               29.560             
\end{verbatim}
 \begin{verbatim}
HF/cc-pVTZ       
       O             -360.518          
       C               -8.791           
       C              165.181            
       H               22.617             
       H               30.044            
       H               30.166             
       H               30.166            
 \end{verbatim}
 \begin{verbatim}
MP2/cc-pVTZ
       O             -284.715           
       C                8.276           
       C              166.742            
       H               22.150            
       H               29.980            
       H               29.780
       H               29.780             
 \end{verbatim}
 \begin{verbatim}
HF/cc-pVQZ       
       O             -360.058          
       C              -13.039           
       C              163.213            
       H               22.510             
       H               29.958             
       H               30.084             
       H               30.084             
 \end{verbatim}
 \begin{verbatim}
MP2/cc-pVQZ
       O             -282.370           
       C                1.640           
       C              163.973           
       H               21.932             
       H               29.790             
       H               29.608             
       H               29.608             
\end{verbatim}

\noindent
   Cholesky threshold = 6
 \begin{verbatim}
HF/cc-pVDZ     
       O             -357.733        
       C                2.771           
       C              171.844            
       H               22.482         
       H               29.600            
       H               29.792             
       H               29.792             
 \end{verbatim}
 \begin{verbatim}
MP2/cc-pVDZ
       O             -253.777          
       C               31.267           
       C              175.828           
       H               22.224             
       H               29.734             
       H               29.560             
       H               29.560             
\end{verbatim}
 \begin{verbatim}
HF/cc-pVTZ 
       O             -360.521         
       C               -8.792           
       C              165.181          
       H               22.617            
       H               30.044           
       H               30.166             
 \end{verbatim}
 \begin{verbatim}
MP2/cc-pVTZ
       O             -284.713           
       C                8.277          
       C              166.743            
       H               22.150            
       H               29.980            
       H               29.780             
       H               29.780             
\end{verbatim}
 \begin{verbatim}
HF/cc-pVQZ   
       O             -360.059          
       C              -13.039          
       C              163.213           
       H               22.510             
       H               29.958            
       H               30.084             
       H               30.084             
 \end{verbatim}
 \begin{verbatim}
MP2/cc-pVQZ
       O             -282.368           
       C                1.641           
       C              163.974            
       H               21.932            
       H               29.790            
       H               29.608             
       H               29.608             
\end{verbatim}

\noindent
without Cholesky decomposition
 \begin{verbatim}
HF/cc-pVDZ
       O             -357.734          
       C                2.771           
       C              171.844            
       H               22.482             
       H               29.600             
       H               29.792             
       H               29.792             
\end{verbatim}
 \begin{verbatim}
MP2/cc-pVDZ
       O             -253.777           
       C               31.267           
       C              175.828            
       H               22.224             
       H               29.734             
       H               29.560             
       H               29.560             
\end{verbatim}
 \begin{verbatim}
HF/cc-pVTZ
       O             -360.522          
       C               -8.792           
       C              165.181            
       H               22.617            
       H               30.044             
       H               30.166             
       H               30.166             
\end{verbatim}
 \begin{verbatim}
MP2/cc-pVTZ
       O             -284.713           
       C                8.277           
       C              166.742           
       H               22.150             
       H               29.980             
       H               29.780             
       H               29.780             
\end{verbatim}
\begin{verbatim}
HF/cc-pVQZ
       O             -360.058          
       C              -13.039           
       C              163.213            
       H               22.510             
       H               29.958             
       H               30.084             
       H               30.084             
\end{verbatim}
 \begin{verbatim}
MP2/cc-pVQZ
       O             -282.367           
       C                1.642          
       C              163.974            
       H               21.932             
       H               29.790             
       H               29.608             
       H               29.608            
\end{verbatim}

\noindent b) ethylene oxide
\\

\noindent Cholesky threshold = 4
\begin{verbatim}
HF/cc-pVDZ
       O              375.060           
       C              166.772            
       C              166.776            
       H               29.529            
       H               29.529            
       H               29.529            
       H               29.529            
\end{verbatim}
\begin{verbatim}
MP2/cc-pVDZ
       O              367.498           
       C              167.561           
       C              167.566            
       H               29.191            
       H               29.191            
       H               29.191           
       H               29.191            
 \end{verbatim} 
\begin{verbatim}
HF/cc-pVTZ
       O              379.303          
       C              158.896            
       C              158.896            
       H               29.884           
       H               29.884            
       H               29.884           
       H               29.884            
\end{verbatim}
\begin{verbatim}
MP2/cc-pVTZ
       O              371.734           
       C              157.589            
       C              157.588            
       H               29.395           
       H               29.395            
       H               29.394            
       H               29.394            
 \end{verbatim} 
 \begin{verbatim}
HF/cc-pVQZ
       O              382.168           
       C              156.391            
       C              156.391            
       H               29.786            
       H               29.786            
       H               29.786            
       H               29.786            
\end{verbatim}
\begin{verbatim}
MP2/cc-pVQZ
       O              375.137           
       C              154.476            
       C              154.477           
       H               29.201            
       H               29.201            
       H               29.201            
       H               29.201            
 \end{verbatim} 

 \noindent   Cholesky threshold = 5
 \begin{verbatim}
HF/cc-pVDZ       
       O              374.997           
       C              166.772            
       C              166.771           
       H               29.529            
       H               29.529            
       H               29.529            
       H               29.529           
 \end{verbatim}
 \begin{verbatim}
MP2/cc-pVDZ
       O              367.449           
       C              167.559            
       C              167.560            
       H               29.192           
       H               29.192           
       H               29.192           
       H               29.192            
\end{verbatim}
 \begin{verbatim}
HF/cc-pVTZ       
       O              379.297           
       C              158.899           
       C              158.899            
       H               29.883            
       H               29.883           
       H               29.883           
       H               29.883            
 \end{verbatim}
 \begin{verbatim}
MP2/cc-pVTZ
       O              371.725           
       C              157.591            
       C              157.591            
       H               29.394            
       H               29.394           
       H               29.394       
       H               29.394            
\end{verbatim}
 \begin{verbatim}
HF/cc-pVQZ       
       O              382.166           
       C              156.390       
       C              156.390           
       H               29.786           
       H               29.786            
       H               29.785            
       H               29.786            
 \end{verbatim}
 \begin{verbatim}
MP2/cc-pVQZ
       O              375.137           
       C              154.479            
       C              154.479            
       H               29.201            
       H               29.201            
       H               29.201           
       H               29.201           
\end{verbatim}

\noindent   Cholesky threshold = 6
 \begin{verbatim}
HF/cc-pVDZ    
       O              374.997       
       C              166.772           
       C              166.772            
       H               29.529           
       H               29.529            
       H               29.529           
       H               29.529           
 \end{verbatim}
 \begin{verbatim}
MP2/cc-pVDZ
       O              367.450           
       C              167.560            
       C              167.560            
       H               29.192            
       H               29.192           
       H               29.192           
       H               29.192            
\end{verbatim}
 \begin{verbatim}
HF/cc-pVTZ    
       O              379.296           
       C              158.900           
       C              158.900           
       H               29.883           
       H               29.883            
       H               29.883            
       H               29.883       
 \end{verbatim}
 \begin{verbatim}
MP2/cc-pVTZ
       O              371.724           
       C              157.593           
       C              157.593            
       H               29.394            
       H               29.394            
       H               29.394            
       H               29.394       
\end{verbatim}
\ \begin{verbatim}
HF/cc-pVQZ    
       O              382.166           
       C              156.390            
       C              156.390           
       H               29.785            
       H               29.785            
       H               29.785            
       H               29.785            
 \end{verbatim}
 \begin{verbatim}
MP2/cc-pVQZ
       O              375.136           
       C              154.479       
       C              154.479       
       H               29.201       
       H               29.201           
       H               29.201           
       H               29.201            
\end{verbatim}

\noindent without Cholesky decomposition
 \begin{verbatim}
HF/cc-pVDZ
       O              374.997           
       C              166.772       
       C              166.772            
       H               29.529       
       H               29.529           
       H               29.529       
       H               29.529           
 \end{verbatim}
 \begin{verbatim}
MP2/cc-pVDZ
       O              367.449           
       C              167.560            
       C              167.560           
       H               29.192       
       H               29.192           
       H               29.192           
       H               29.192       
 \end{verbatim}
 \begin{verbatim}
HF/cc-pVTZ
       O              379.296   
       C              158.900   
       C              158.900            
       H               29.883            
       H               29.883            
       H               29.883            
       H               29.883            
 \end{verbatim}
 \begin{verbatim}
MP2/cc-pVTZ
       O              371.724           
       C              157.593            
       C              157.593            
       H               29.394            
       H               29.394       
       H               29.394            
       H               29.394            
  \end{verbatim}
 \begin{verbatim}
HF/cc-pvQZ
       O              382.165   
       C              156.390            
       C              156.390           
       H               29.785           
       H               29.785           
       H               29.785            
       H               29.785            
 \end{verbatim}
 \begin{verbatim}
MP2/cc-pVQZ
       O              375.136       
       C              154.480           
       C              154.480           
       H               29.201            
       H               29.201           
       H               29.201       
       H               29.201            
 \end{verbatim}
\noindent

\noindent c) vinyl alcohol
\\

\noindent Cholesky threshold = 4
\begin{verbatim}
HF/cc-pVDZ
       O              245.947            
       C               47.407           
       C              114.048            
       H               28.609           
       H               25.339             
       H               27.707             
       H               27.581             
\end{verbatim}
\begin{verbatim}
MP2/cc-pVDZ
       O              251.240            
       C               64.790           
       C              128.452  
       H               28.486            
       H               25.286             
       H               27.773             
       H               27.618            
 \end{verbatim} 
\begin{verbatim}
HF/cc-pVTZ
       O              238.962           
       C               36.344           
       C              106.936           
       H               28.183            
       H               25.599             
       H               28.006            
       H               27.953             
\end{verbatim}
\begin{verbatim}
MP2/cc-pVTZ
       O              241.116            
       C               45.977       
       C              115.690            
       H               27.943           
       H               25.345            
       H               27.828             
       H               27.772            
 \end{verbatim} 
 \begin{verbatim}
HF/cc-pVQZ
       O              238.265            
       C               33.201           
       C              104.227           
       H               28.005           
       H               25.447           
       H               27.927            
       H               27.859             
\end{verbatim}
\begin{verbatim}
MP2/cc-pVQZ
       O              240.150            
       C               41.595           
       C              111.766           
       H               27.618            
       H               25.089            
       H               27.632           
       H               27.575            
 \end{verbatim} 

 \noindent   Cholesky threshold = 5
 \begin{verbatim}
HF/cc-pVDZ 
       O              245.934           
       C               47.417           
       C              114.040           
       H               28.610           
       H               25.339            
       H               27.707            
       H               27.581            
 \end{verbatim}
 \begin{verbatim}
MP2/cc-pVDZ
       O              251.235            
       C               64.804           
       C              128.451       
       H               28.487            
       H               25.286             
       H               27.774            
       H               27.619           
\end{verbatim}
 \begin{verbatim}
HF/cc-pVTZ     
       O              238.962           
       C               36.344           
       C              106.933            
       H               28.183           
       H               25.599            
       H               28.006            
       H               27.953             
 \end{verbatim}
 \begin{verbatim}
MP2/cc-pVTZ
       O              241.117           
       C               45.978       
       C              115.691            
       H               27.943            
       H               25.346            
       H               27.828             
       H               27.772            
\end{verbatim}
 \begin{verbatim}
HF/cc-pVQZ  
       O              238.244            
       C               33.089           
       C              104.143           
       H               28.004           
       H               25.447            
       H               27.926             
       H               27.858            
 \end{verbatim}
 \begin{verbatim}
MP2/cc-pVQZ
       O              240.134           
       C               41.513       
       C              111.705           
       H               27.618            
       H               25.089             
       H               27.631            
       H               27.575            
\end{verbatim}

\noindent   Cholesky threshold = 6
 \begin{verbatim}
HF/cc-pVDZ    
       O              245.934            
       C               47.416           
       C              114.038            
       H               28.610           
       H               25.339             
       H               27.707             
       H               27.581             
 \end{verbatim}
 \begin{verbatim}
MP2/cc-pVDZ
       O              251.236            
       C               64.804           
       C              128.450            
       H               28.486            
       H               25.287             
       H               27.774             
       H               27.619            
\end{verbatim}
 \begin{verbatim}
HF/cc-pVTZ  
       O              238.960            
       C               36.344       
       C              106.934            
       H               28.183           
       H               25.599             
       H               28.006             
       H               27.953             
 \end{verbatim}
 \begin{verbatim}
MP2/cc-pVTZ
       O              241.116            
       C               45.978           
       C              115.693            
       H               27.943            
       H               25.345             
       H               27.828             
       H               27.772             
\end{verbatim}
 \begin{verbatim}
HF/cc-pVQZ   
       O              238.244            
       C               33.089           
       C              104.142            
       H               28.004            
       H               25.447             
       H               27.926             
       H               27.858             
 \end{verbatim}
 \begin{verbatim}
MP2/cc-pVQZ
       O              240.134            
       C               41.514           
       C              111.706            
       H               27.618            
       H               25.089             
       H               27.631            
       H               27.575            
\end{verbatim}

\noindent without Cholesky decomposition
 \begin{verbatim}
HF/cc-pVDZ
       O              245.933            
       C               47.416           
       C              114.038            
       H               28.610            
       H               25.339             
       H               27.707             
       H               27.581             
 \end{verbatim}
 \begin{verbatim}
MP2/cc-pVDZ
       O              251.236            
       C               64.804           
       C              128.450            
       H               28.486            
       H               25.287             
       H               27.774             
       H               27.619             
 \end{verbatim}
 \begin{verbatim}
HF/cc-pVTZ
       O              238.960            
       C               36.343           
       C              106.934            
       H               28.183            
       H               25.599             
       H               28.006             
       H               27.953             
 \end{verbatim}
 \begin{verbatim}
MP2/cc-pVTZ
       O              241.116            
       C               45.978           
       C              115.693            
       H               27.943            
       H               25.345             
       H               27.828             
       H               27.772             
  \end{verbatim}
 \begin{verbatim}
HF/cc-pvQZ
       O              238.244            
       C               33.089           
       C              104.142            
       H               28.004            
       H               25.447             
       H               27.926             
       H               27.858             
 \end{verbatim}
 \begin{verbatim}
MP2/cc-pVQZ
       O              240.134            
       C               41.514           
       C              111.706            
       H               27.618            
       H               25.089             
       H               27.631            
       H               27.575            
 \end{verbatim}
\noindent

\noindent

\section{Geometries used in the CD based GIAO-MP2 calculations}

In this section, we document the geometries used in the CD based GIAO-MP2 calculations reported in section IV.C of the main document. All coordinates are given in \AA{}ngstr\"om.\\

coronene
\begin{verbatim}
c     -0.515149   -1.337548    0.000025
c      0.900713   -1.114666    0.000032
c     -1.415886   -0.222738    0.000019
c      1.415903    0.222760    0.000031
c     -0.900696    1.114688    0.000018
c      0.515165    1.337570    0.000025
c     -1.030023   -2.674617    0.000026
c      1.801114   -2.229165    0.000038
c     -2.831321   -0.445493    0.000013
c      2.831338    0.445515    0.000038
c     -1.801097    2.229187    0.000012
c      1.030040    2.674639    0.000024
c     -0.107792   -3.767248    0.000032
c     -2.446919   -2.866601    0.000020
c      1.259004   -3.552179    0.000038
c      3.208445   -1.976781    0.000044
c     -3.316572   -1.790447    0.000013
c     -3.705806    0.685730    0.000006
c      3.705822   -0.685708    0.000044
c      3.316588    1.790469    0.000037
c     -3.208428    1.976803    0.000006
c     -1.258987    3.552201    0.000011
c      2.446935    2.866624    0.000030
c      0.107809    3.767270    0.000017
h     -0.507491   -4.793413    0.000033
h     -2.838757   -3.895787    0.000020
h      1.954169   -4.406366    0.000043
h      3.897339   -2.835959    0.000049
h     -4.405014   -1.957667    0.000009
h     -4.793104    0.511169    0.000002
h      4.793120   -0.511146    0.000049
h      4.405030    1.957689    0.000042
h     -3.897322    2.835981    0.000000
h     -1.954152    4.406388    0.000006
h      2.838774    3.895810    0.000030
h      0.507508    4.793435    0.000017
\end{verbatim}

hexabenzocoronene
\begin{verbatim}
c    2.3455393628573100       -1.3541977825405300        0.0000000000000000
c    2.3455393628573100        1.3541977825405300        0.0000000000000000
c    0.0000000000000000        2.7083955650810698        0.0000000000000000
c   -2.3455393628573100        1.3541977825405300        0.0000000000000000
c   -2.3455393628573100       -1.3541977825405300        0.0000000000000000
c    0.0000000000000000       -2.7083955650810698        0.0000000000000000
c    4.7200845444155899       -2.7251420823161299        0.0000000000000000
c    4.7200845444155899        2.7251420823161299        0.0000000000000000
c    0.0000000000000000        5.4502841646322597        0.0000000000000000
c   -4.7200845444155899        2.7251420823161299        0.0000000000000000
c   -4.7200845444155899       -2.7251420823161299        0.0000000000000000
c    0.0000000000000000       -5.4502841646322597        0.0000000000000000
c    7.0769289511666997       -1.3842901466443300        0.0000000000000000
c    4.7372949087858300        5.4366551791657596        0.0000000000000000
c   -2.3396340423808701        6.8209453258100901        0.0000000000000000
c   -7.0769289511666997        1.3842901466443400        0.0000000000000000
c   -4.7372949087858300       -5.4366551791657498        0.0000000000000000
c    2.3396340423808599       -6.8209453258100901        0.0000000000000000
c    4.7372949087858300       -5.4366551791657596        0.0000000000000000
c   -2.3396340423808701       -6.8209453258100901        0.0000000000000000
c   -7.0769289511666997       -1.3842901466443400        0.0000000000000000
c   -4.7372949087858300        5.4366551791657498        0.0000000000000000
c    2.3396340423808599        6.8209453258100901        0.0000000000000000
c    7.0769289511666997        1.3842901466443300        0.0000000000000000
c    9.3595921381599005       -2.7665262303351099        0.0000000000000000
c    7.0756780647861603        6.7223814455400301        0.0000000000000000
c   -2.2839140733737402        9.4889076758751401        0.0000000000000000
c   -9.3595921381599005        2.7665262303351099        0.0000000000000000
c   -7.0756780647861603       -6.7223814455400301        0.0000000000000000
c    2.2839140733737402       -9.4889076758751401        0.0000000000000000
c    7.0756780647861603       -6.7223814455400301        0.0000000000000000
c   -2.2839140733737402       -9.4889076758751401        0.0000000000000000
c   -9.3595921381599005       -2.7665262303351099        0.0000000000000000
c   -7.0756780647861603        6.7223814455400301        0.0000000000000000
c    2.2839140733737402        9.4889076758751401        0.0000000000000000
c    9.3595921381599005        2.7665262303351099        0.0000000000000000
c    9.3618315723958805       -5.4050559784307000        0.0000000000000000
c    9.3618315723958805        5.4050559784306902        0.0000000000000000
c    0.0000000000000000        10.810111956861389        0.0000000000000000
c   -9.3618315723958805        5.4050559784307000        0.0000000000000000
c   -9.3618315723958805       -5.4050559784307000        0.0000000000000000
c    0.0000000000000000       -10.810111956861389        0.0000000000000000
h    11.180923626803910        1.7743617216381000        0.0000000000000000
h    4.0538194869606698        10.570144759404871        0.0000000000000000
h   -7.1271041398432402        8.7957830377667801        0.0000000000000000
h   -11.180923626803910       -1.7743617216380900        0.0000000000000000
h   -4.0538194869606698       -10.570144759404871        0.0000000000000000
h    7.1271041398432304       -8.7957830377667801        0.0000000000000000
h    4.0538194869606698       -10.570144759404871        0.0000000000000000
h   -7.1271041398432402       -8.7957830377667801        0.0000000000000000
h   -11.180923626803910        1.7743617216380900        0.0000000000000000
h   -4.0538194869606698        10.570144759404871        0.0000000000000000
h    7.1271041398432304        8.7957830377667801        0.0000000000000000
h    11.180923626803910       -1.7743617216381000        0.0000000000000000
h    11.162923234837629       -6.4449167345766298        0.0000000000000000
h    11.162923234837629        6.4449167345766298        0.0000000000000000
h    0.0000000000000000        12.889833469153270        0.0000000000000000
h   -11.162923234837629        6.4449167345766298        0.0000000000000000
h   -11.162923234837629       -6.4449167345766298        0.0000000000000000
h    0.0000000000000000       -12.889833469153260        0.0000000000000000
\end{verbatim}

B$_4$tBu$_4$
\begin{verbatim}
b    1.1690489615997499        1.1690489615997499        1.1690489615997499
b   -1.1690489615997499        1.1690489615997499       -1.1690489615997499
b   -1.1690489615997499       -1.1690489615997499        1.1690489615997499
b    1.1690489615997499       -1.1690489615997499       -1.1690489615997499
c    2.9564738520445002        2.9564738520445002        2.9564738520445002
c    2.4540179688007102        5.8215137664078203        2.4540179688007102
c    2.4540179688007102        2.4540179688007102        5.8215137664078203
c    5.8215137664078203        2.4540179688007102        2.4540179688007102
h    1.0308755144433499        6.1556950098414100        1.0308755144433499
h    1.8122174063511900        6.7903696527766204        4.1607412464994997
h    4.1607412464994997        6.7903696527766204        1.8122174063511900
h    1.0308755144433499        1.0308755144433499        6.1556950098414100
h    4.1607412464994997        1.8122174063511900        6.7903696527766204
h    1.8122174063511900        4.1607412464994997        6.7903696527766204
h    6.1556950098414100        1.0308755144433499        1.0308755144433499
h    6.7903696527766204        4.1607412464994997        1.8122174063511900
h    6.7903696527766204        1.8122174063511900        4.1607412464994997
c   -2.9564738520445002        2.9564738520445002       -2.9564738520445002
c   -2.9564738520445002       -2.9564738520445002        2.9564738520445002
c    2.9564738520445002       -2.9564738520445002       -2.9564738520445002
c   -5.8215137664078203        2.4540179688007102       -2.4540179688007102
c   -2.4540179688007102       -5.8215137664078203        2.4540179688007102
c    5.8215137664078203       -2.4540179688007102       -2.4540179688007102
c    2.4540179688007102       -2.4540179688007102       -5.8215137664078203
c   -2.4540179688007102        5.8215137664078203       -2.4540179688007102
c   -2.4540179688007102        2.4540179688007102       -5.8215137664078203
c   -5.8215137664078203       -2.4540179688007102        2.4540179688007102
c   -2.4540179688007102       -2.4540179688007102        5.8215137664078203
c    2.4540179688007102       -5.8215137664078203       -2.4540179688007102
h   -6.1556950098414100        1.0308755144433499       -1.0308755144433499
h   -1.0308755144433499       -6.1556950098414100        1.0308755144433499
h    6.1556950098414100       -1.0308755144433499       -1.0308755144433499
h    1.0308755144433499       -1.0308755144433499       -6.1556950098414100
h   -1.0308755144433499        6.1556950098414100       -1.0308755144433499
h   -1.0308755144433499        1.0308755144433499       -6.1556950098414100
h   -6.1556950098414100       -1.0308755144433499        1.0308755144433499
h   -1.0308755144433499       -1.0308755144433499        6.1556950098414100
h    1.0308755144433499       -6.1556950098414100       -1.0308755144433499
h   -6.7903696527766204        1.8122174063511900       -4.1607412464994997
h   -1.8122174063511900       -6.7903696527766204        4.1607412464994997
h    6.7903696527766204       -1.8122174063511900       -4.1607412464994997
h   -4.1607412464994997       -6.7903696527766204        1.8122174063511900
h    4.1607412464994997       -1.8122174063511900       -6.7903696527766204
h   -4.1607412464994997        6.7903696527766204       -1.8122174063511900
h   -1.8122174063511900        4.1607412464994997       -6.7903696527766204
h    6.7903696527766204       -4.1607412464994997       -1.8122174063511900
h   -6.7903696527766204       -4.1607412464994997        1.8122174063511900
h   -4.1607412464994997       -1.8122174063511900        6.7903696527766204
h   -4.1607412464994997        1.8122174063511900       -6.7903696527766204
h    4.1607412464994997       -6.7903696527766204       -1.8122174063511900
h    1.8122174063511900       -4.1607412464994997       -6.7903696527766204
h   -6.7903696527766204        4.1607412464994997       -1.8122174063511900
h   -1.8122174063511900       -4.1607412464994997        6.7903696527766204
h   -6.7903696527766204       -1.8122174063511900        4.1607412464994997
h   -1.8122174063511900        6.7903696527766204       -4.1607412464994997
h    1.8122174063511900       -6.7903696527766204       -4.1607412464994997
\end{verbatim}

Al$_4$Cp$_4$
\begin{verbatim}
al  -1.7848909324915401        1.7848909324915401       -1.8016806976698900
al  -1.7848909324915401       -1.7848909324915401        1.8016806976698900
al   1.7848909324915401       -1.7848909324915401       -1.8016806976698900
c    2.1327918704218400        5.2068227625738501        4.6011710656555103
c    3.7810413517918202        5.6797991806534904        2.5322310236310801
c    5.6797991806534904        3.7810413517918202        2.5322310236310801
c    5.2068227625738501        2.1327918704218400        4.6011710656555103
c    3.0170803931520300        3.0170803931520300        5.8809048963372996
h    0.45717289155772001       6.3025686837033401        5.0909211513568202
h    3.5907636308872899        7.2017409444757101        1.1554131263342100
h    7.2017409444757101        3.5907636308872899        1.1554131263342100
h    6.3025686837033401        0.45717289155772001       5.0909211513568202
h    2.1341569998567200        2.1341569998567200        7.5210033038361601
al   1.7848909324915401        1.7848909324915401        1.8016806976698900
c   -5.2068227625738501        2.1327918704218400       -4.6011710656555103
c   -2.1327918704218400       -5.2068227625738501        4.6011710656555103
c    5.2068227625738501       -2.1327918704218400       -4.6011710656555103
c    2.1327918704218400       -5.2068227625738501       -4.6011710656555103
c   -5.2068227625738501       -2.1327918704218400        4.6011710656555103
c   -2.1327918704218400        5.2068227625738501       -4.6011710656555103
c   -5.6797991806534904        3.7810413517918202       -2.5322310236310801
c   -3.7810413517918202       -5.6797991806534904        2.5322310236310801
c    5.6797991806534904       -3.7810413517918202       -2.5322310236310801
c    3.7810413517918202       -5.6797991806534904       -2.5322310236310801
c   -5.6797991806534904       -3.7810413517918202        2.5322310236310801
c   -3.7810413517918202        5.6797991806534904       -2.5322310236310801
c   -3.0170803931520300        3.0170803931520300       -5.8809048963372996
c   -3.0170803931520300       -3.0170803931520300        5.8809048963372996
c    3.0170803931520300       -3.0170803931520300       -5.8809048963372996
h   -6.3025686837033401        0.45717289155772001      -5.0909211513568202
h   -0.45717289155772001      -6.3025686837033401        5.0909211513568202
h    6.3025686837033401       -0.45717289155772001      -5.0909211513568202
h    0.45717289155772001      -6.3025686837033401       -5.0909211513568202
h   -6.3025686837033401       -0.45717289155772001       5.0909211513568202
h   -0.45717289155772001       6.3025686837033401       -5.0909211513568202
h   -7.2017409444757101        3.5907636308872899       -1.1554131263342100
h   -3.5907636308872899       -7.2017409444757101        1.1554131263342100
h    7.2017409444757101       -3.5907636308872899       -1.1554131263342100
h    3.5907636308872899       -7.2017409444757101       -1.1554131263342100
h   -7.2017409444757101       -3.5907636308872899        1.1554131263342100
h   -3.5907636308872899        7.2017409444757101       -1.1554131263342100
h   -2.1341569998567200        2.1341569998567200       -7.5210033038361601
h   -2.1341569998567200       -2.1341569998567200        7.5210033038361601
h    2.1341569998567200       -2.1341569998567200       -7.5210033038361601
\end{verbatim}

C$_{60}$
\begin{verbatim}
c    0.0000000000000000       -2.3324791571039798        6.3117292845539996
c    2.2183195014863699       -0.72077569857047996       6.3117292845539996
c    1.3709968498253000        1.8870152771224700        6.3117292845539996
c   -1.3709968498253000        1.8870152771224700        6.3117292845539996
c   -2.2183195014863699       -0.72077569857047996       6.3117292845539996
c    4.9234635270203597        4.4440870275860798        1.1348933734771900
c    2.7051440255339898        6.0557904861195704        1.1348933734771900
c    1.3341471757086900        5.6103266061380701        3.4673725305811698
c    2.7051440255339898        3.7233113290155999        4.9089239277221397
c    4.9234635270203597        3.0025356304451098        3.4673725305811698
c   -2.7051440255339898        6.0557904861195704        1.1348933734771900
c   -4.9234635270203597        4.4440870275860798        1.1348933734771900
c   -4.9234635270203597        3.0025356304451098        3.4673725305811698
c   -2.7051440255339898        3.7233113290155999        4.9089239277221397
c   -1.3341471757086900        5.6103266061380701        3.4673725305811698
c   -6.5953344792640802       -0.70140267841592996       1.1348933734771900
c   -5.7480118276030101       -3.3091936541088902        1.1348933734771900
c   -4.3770149777777103       -3.7546575340903900        3.4673725305811698
c   -4.3770149777777103       -1.4221783769864200        4.9089239277221397
c   -5.7480118276030101        0.46483690013605999       3.4673725305811698
c   -1.3709968498253000       -6.4892811811808304        1.1348933734771900
c    1.3709968498253000       -6.4892811811808304        1.1348933734771900
c    2.2183195014863699       -5.3230416026288401        3.4673725305811698
c    0.0000000000000000       -4.6022659040583598        4.9089239277221397
c   -2.2183195014863699       -5.3230416026288401        3.4673725305811698
c    5.7480118276030101       -3.3091936541088902        1.1348933734771900
c    6.5953344792640802       -0.70140267841592996       1.1348933734771900
c    5.7480118276030101        0.46483690013605999       3.4673725305811698
c    4.3770149777777103       -1.4221783769864200        4.9089239277221397
c    4.3770149777777103       -3.7546575340903900        3.4673725305811698
c    5.7480118276030101       -0.46483690013605999      -3.4673725305811698
c    4.3770149777777103        1.4221783769864200       -4.9089239277221397
c    4.3770149777777103        3.7546575340903900       -3.4673725305811698
c    5.7480118276030101        3.3091936541088902       -1.1348933734771900
c    6.5953344792640802        0.70140267841592996      -1.1348933734771900
c    2.2183195014863699        5.3230416026288401       -3.4673725305811698
c    0.0000000000000000        4.6022659040583598       -4.9089239277221397
c   -2.2183195014863699        5.3230416026288401       -3.4673725305811698
c   -1.3709968498253000        6.4892811811808304       -1.1348933734771900
c    1.3709968498253000        6.4892811811808304       -1.1348933734771900
c   -4.3770149777777103        3.7546575340903900       -3.4673725305811698
c   -4.3770149777777103        1.4221783769864200       -4.9089239277221397
c   -5.7480118276030101       -0.46483690013605999      -3.4673725305811698
c   -6.5953344792640802        0.70140267841592996      -1.1348933734771900
c   -5.7480118276030101        3.3091936541088902       -1.1348933734771900
c   -4.9234635270203597       -3.0025356304451098       -3.4673725305811698
c   -2.7051440255339898       -3.7233113290155999       -4.9089239277221397
c   -1.3341471757086900       -5.6103266061380701       -3.4673725305811698
c   -2.7051440255339898       -6.0557904861195704       -1.1348933734771900
c   -4.9234635270203597       -4.4440870275860798       -1.1348933734771900
c    1.3341471757086900       -5.6103266061380701       -3.4673725305811698
c    2.7051440255339898       -3.7233113290155999       -4.9089239277221397
c    4.9234635270203597       -3.0025356304451098       -3.4673725305811698
c    4.9234635270203597       -4.4440870275860798       -1.1348933734771900
c    2.7051440255339898       -6.0557904861195704       -1.1348933734771900
c    1.3709968498253000       -1.8870152771224700       -6.3117292845539996
c   -1.3709968498253000       -1.8870152771224700       -6.3117292845539996
c   -2.2183195014863699        0.72077569857047996      -6.3117292845539996
c    0.0000000000000000        2.3324791571039798       -6.3117292845539996
c    2.2183195014863699        0.72077569857047996      -6.3117292845539996
\end{verbatim}

tweezer host-guest complex
\begin{verbatim}
n      5.323376    8.894292   15.683438
h      6.008335    5.939395   14.844914
h      6.728273    6.998069   12.237493
h      6.392142    9.727046   13.165319
c      6.262998    8.852781   16.387144
h      6.051308    5.626441   20.711084
h      5.997106    6.772760   23.401417
h      5.617173    9.514511   22.461966
h      6.237570    8.810903   19.088449
h      6.792653   12.530158   13.779448
h      6.146629   12.155297   16.583965
h      5.919504   12.377819   21.940614
h      5.887650   12.104602   19.050723
c      9.377810    6.257590   14.164567
c      9.926215    5.682276   15.310076
h     11.016157    5.625785   15.456396
c      9.043718    5.168591   16.287568
h      9.453991    4.700972   17.193732
c      7.653156    5.254527   16.118705
h      6.982243    4.857714   16.895181
c      7.098333    5.850326   14.964286
c      7.971701    6.334060   13.992072
c      7.734770    7.093126   12.681579
c     10.002059    6.977921   12.959177
h     11.070414    6.772643   12.766310
c      8.962299    6.579807   11.862203
h      9.107111    7.132495   10.912229
h      8.930308    5.487680   11.674779
c      9.615596    8.451991   13.149839
c     10.361156    9.570561   13.551983
h     11.450394    9.514603   13.705777
c      9.647310   10.759931   13.767308
c      8.238262   10.815569   13.615030
c      7.489468    9.695039   13.230985
c      8.207833    8.518363   12.984638
c      9.822090    8.807060   17.515059
h     10.830319    8.810852   17.079487
c      8.704796    8.801295   16.680516
h      8.825175    8.794557   15.590961
c      7.409615    8.815193   17.246689
n     11.744423    8.894566   20.484590
c      7.689681    6.257865   22.003064
c      7.141251    5.682810   20.857440
c      8.023735    5.169353   19.879815
h      7.613453    4.702048   18.973497
c      9.414299    5.255172   20.048710
h     10.085188    4.858583   19.272100
c      9.969149    5.850705   21.203255
h     11.059154    5.939670   21.322668
c      9.095797    6.334290   22.175558
c      9.332758    7.093158   23.486168
h     10.339256    6.998017   23.930233
C      7.065467    6.978023   23.208564
C      7.451967    8.452089   23.018039
C      6.706388    9.570635   22.615891
C      7.420209   10.760032   22.400636
C      8.829257   10.815687   22.552950
C      9.578069    9.695153   22.936969
h     10.675398    9.727174   23.002606
c      8.859721    8.518453   23.183282
c      7.245799    8.807116   18.652873
c      8.363097    8.801538   19.487410
h      8.242844    8.795009   20.576979
c      9.658272    8.815336   18.921232
c     10.804862    8.853097   19.780803
c      8.105225    6.579754   24.305478
h      7.960428    7.132316   25.255527
h      8.137203    5.487602   24.492756
c      7.801513   12.203787   14.087750
c     10.077315   12.123644   14.325068
h     11.148005   12.377749   14.227276
c      9.032288   13.039536   13.608919
h      9.029798   14.077357   13.998843
h      9.146406   13.040106   12.506286
c      9.500504   12.159341   15.745115
c     10.085405   12.124874   16.993644
h     11.179915   12.104696   17.117201
c      7.813286   12.136269   18.007769
c      7.242342   12.157645   16.694460
c      8.075197   12.193467   15.595946
c      9.265995   12.203911   22.080216
h     10.274844   12.530296   22.388538
c      6.990201   12.123729   21.842854
c      8.035195   13.039648   22.559005
h      8.037675   14.077461   22.169059
h      7.921053   13.040238   23.661635
c      7.567042   12.159396   20.422822
c      6.982154   12.124844   19.174292
c      9.254279   12.136280   18.160177
c      9.825212   12.157730   19.473495
h     10.920926   12.155458   19.583996
c      8.992346   12.193556   20.572007
\end{verbatim}

\section{{dzp}, {tzp}, and {tz2p} basis sets used in the CD based GIAO-MP2 calculations}

In this section, we document the dzp, tzp, and tz2p basis sets used in the representative CD based GIAO-MP2 calculations reported in section IV.C of the main document. The basis sets are given in the format required
for calculations using the CFOUR program package (see www.cfour.de for a detailed description of the format).\\

\subsection{dzp basis set}

\begin{verbatim}
H:dzp
double zeta (4s primitive set) plus polarization (Ahlrichs basis)

  2
    0    1
    2    1
    4    1

    13.0107010     1.9622572      .4445380      .1219496

  .0196822   .0000000
  .1379652   .0000000
  .4783193   .0000000
  .0000000  1.0000000

      .8000000

 1.0000000

B:dzp
double polarization (8s4p primitive) plus polarization (Ahlrichs basis)

  6
    0    0    0    1    1    2
    1    2    1    1    1    1
    5    2    1    3    1    1

  2410.2061000   361.8699200    82.3085930    23.1099420     7.2517259

  .0017550   .0000000   .0000000   .0000000
  .0134364   .0000000   .0000000   .0000000
  .0663690   .0000000   .0000000   .0000000
  .2302530   .0000000   .0000000   .0000000
  .5149821   .0000000   .0000000   .0000000

     2.3666469      .3598708

 1.0000000   .0000000   .0000000
  .0000000  1.0000000   .0000000

      .1116289

 1.0000000

     6.0017073     1.2401097      .3366802

 -.0355453   .0000000
 -.1983450   .0000000
 -.5047873   .0000000

      .0955909

 1.0000000

      .5000000

 1.0000000

C:dzp
double zeta plus polarization (Ahlrichs basis)

  6
    0    0    0    1    1    2
    1    2    1    1    1    1
    5    2    1    3    1    1

  3623.8613000   544.0462100   123.7433800    34.7632090    10.9333330

  .0016339   .0000000   .0000000   .0000000
  .0125217   .0000000   .0000000   .0000000
  .0621139   .0000000   .0000000   .0000000
  .2181773   .0000000   .0000000   .0000000
  .4980043   .0000000   .0000000   .0000000

     3.5744765      .5748325

 1.0000000   .0000000   .0000000
  .0000000  1.0000000   .0000000

      .1730364

 1.0000000

     9.4432819     2.0017986      .5462972

  .0378955   .0000000
  .2081818   .0000000
  .5047417   .0000000

     .1520268

 1.0000000

     .8000000

 1.0000000

N:dzp
double zeta (8s4p primitive set) plus polarization (Ahlrichs basis)

  6
    0    0    0    1    1    2
    1    2    1    1    1    1
    5    2    1    3    1    1

  5071.9892000   761.4179100   173.1841800    48.6703900    15.3314480

  .0017067   .0000000   .0000000   .0000000
  .0130877   .0000000   .0000000   .0000000
  .0650983   .0000000   .0000000   .0000000
  .2305160   .0000000   .0000000   .0000000
  .5330047   .0000000   .0000000   .0000000

     5.0186710      .8355477

  1.0000000   .0000000   .0000000
   .0000000  1.0000000   .0000000

      .2462681

  1.0000000   .0000000   .0000000

    13.5507220     2.9178682      .7983125

  .0406308   .0000000
  .2208580   .0000000
  .5186014   .0000000

      .2190013

 1.0000000

     1.000000

 1.0000000

AL:dzp
double zeta basis plus polarization (Ahlrichs basis)

  7
    0    0    0    1    1    1    2
    1    2    3    1    2    1    1
    5    3    3    4    2    1    1

 32386.3810000  4858.4056000  1105.9177000   313.3923900   102.6549500

  .0006266   .0000000   .0000000   .0000000   .0000000   .0000000
  .0048376   .0000000   .0000000   .0000000   .0000000   .0000000
  .0247057   .0000000   .0000000   .0000000   .0000000   .0000000
  .0942671   .0000000   .0000000   .0000000   .0000000   .0000000
  .2565490   .0000000   .0000000   .0000000   .0000000   .0000000

    37.4098730    14.4578780     3.2405266

  .4546558   .0000000   .0000000   .0000000   .0000000
  .2958714   .0000000   .0000000   .0000000   .0000000
  .0000000  1.0000000   .0000000   .0000000   .0000000

     1.1620812      .1769117      .0654140

 1.0000000   .0000000   .0000000
  .0000000  1.0000000   .0000000
  .0000000   .0000000  1.0000000

   145.6757800    33.8475080    10.4135470     3.5311030

  .0105725   .0000000   .0000000   .0000000
  .0730995   .0000000   .0000000   .0000000
  .2579905   .0000000   .0000000   .0000000
  .4753818   .0000000   .0000000   .0000000

     1.2051750     .2718885

 1.0000000   .0000000   .0000000
  .0000000  1.0000000   .0000000

     .0714647

 1.0000000

      .3000000

 1.0000000

\end{verbatim}

 \subsection{tzp basis set}

\begin{verbatim}
H:tzp
triple zeta plus double polariation (Ahlrichs basis)

  3
    0    0    1
    1    2    1
    3    2    1

    34.0613410     5.1235746     1.1646626

  .0060252
  .0450211
  .2018973

   .3272304      .1030724

 1.0000000   .0000000
  .0000000  1.0000000

   .800000

 1.0000000

C:tzp
triple zeta (9s5p primitive set) plus double polarization (Ahlrichs basis)

 6
    0    0    0    1    1    2
    1    2    2    1    2    1
    5    2    2    3    2    1

  7156.1744000  1073.4735000   244.3065600    69.0834620    22.3409630

  .0006699   .0000000   .0000000   .0000000   .0000000
  .0051710   .0000000   .0000000   .0000000   .0000000
  .0264024   .0000000   .0000000   .0000000   .0000000
  .1009319   .0000000   .0000000   .0000000   .0000000
  .2802866   .0000000   .0000000   .0000000   .0000000

     7.8524164     2.8791374

 1.0000000   .0000000   .0000000   .0000000
 0.0000000  1.0000000   .0000000   .0000000

      .5194815      .1593198

 1.0000000   .0000000   .0000000   .0000000
 0.0000000  1.0000000   .0000000   .0000000

    18.7360050     4.1363791     1.2005194

  .0140064   .0000000   .0000000
  .0868683   .0000000   .0000000
  .2899842   .0000000   .0000000

      .3834848      .1212946

  1.0000000   .0000000
   .0000000  1.0000000

      .8000

 1.0000000

N:tzp
triple zeta (9s5p primitive set) plus polarization (Ahlrichs basis)

  6
    0    0    0    1    1    2
    1    2    2    1    2    1
    5    2    2    3    2    1

  9810.2727000  1471.5960000   334.9146600    94.7199660    30.6694460

  .0006755   .0000000   .0000000   .0000000   .0000000
  .0052154   .0000000   .0000000   .0000000   .0000000
  .0266464   .0000000   .0000000   .0000000   .0000000
  .1020819   .0000000   .0000000   .0000000   .0000000
  .2843982   .0000000   .0000000   .0000000   .0000000

    10.8195540     3.9873030

 1.0000000   .0000000   .0000000   .0000000
  .0000000  1.0000000   .0000000   .0000000

      .7448334      .2247018

 1.0000000   .0000000   .0000000   .0000000
  .0000000  1.0000000   .0000000   .0000000

    26.6716420     5.9564688     1.7441906

  .0146417   .0000000   .0000000
  .0916834   .0000000   .0000000
  .2984215   .0000000   .0000000

      .5563397      .1731605

 1.0000000   .0000000
  .0000000  1.0000000

     1.000000

 1.0000000

\end{verbatim}
 \subsection{tz2p basis set}
\begin{verbatim}
H:tz2p
triple zeta plus double polariation (Ahlrichs basis)

  4
    0    0    1    1
    1    2    1    1
    3    2    1    1

    34.0613410     5.1235746     1.1646626

  .0060252
  .0450211
  .2018973

   .3272304      .1030724

 1.0000000   .0000000
  .0000000  1.0000000

      .4600000

 1.0000000

      1.39000000

 1.0000000

B:tz2p
triple zeta (9s5p primitives) plus double polarization (Ahlrichs basis)

  7
    0    0    0    1    1    2    2
    1    2    2    1    2    1    1
    5    2    2    3    2    1    1

  4896.2210000   734.4699800   167.1525300    47.2546880    15.2538150

  .0006649   .0000000   .0000000   .0000000   .0000000
  .0051312   .0000000   .0000000   .0000000   .0000000
  .0261714   .0000000   .0000000   .0000000   .0000000
  .0997066   .0000000   .0000000   .0000000   .0000000
  .2753155   .0000000   .0000000   .0000000   .0000000

     5.3325783     1.9406529

 1.0000000   .0000000   .0000000   .0000000
  .0000000  1.0000000   .0000000   .0000000

      .3313052      .1040601

 1.0000000   .0000000
  .0000000  1.0000000

    12.0543230     2.6121298      .7469676

 -.0134352   .0000000   .0000000
 -.0819531   .0000000   .0000000
 -.2840766   .0000000   .0000000

      .2387489      .0772200

 1.0000000   .0000000
  .0000000  1.0000000

      .2900000

 1.0000000

      .8700000

 1.0000000

C:tz2p
triple zeta (9s5p primitive set) plus double polarization (Ahlrichs basis)

  7
    0    0    0    1    1    2    2
    1    2    2    1    2    1    1
    5    2    2    3    2    1    1

  7156.1744000  1073.4735000   244.3065600    69.0834620    22.3409630

  .0006699   .0000000   .0000000   .0000000   .0000000
  .0051710   .0000000   .0000000   .0000000   .0000000
  .0264024   .0000000   .0000000   .0000000   .0000000
  .1009319   .0000000   .0000000   .0000000   .0000000
  .2802866   .0000000   .0000000   .0000000   .0000000

     7.8524164     2.8791374

 1.0000000   .0000000   .0000000   .0000000
 0.0000000  1.0000000   .0000000   .0000000

      .5194815      .1593198

 1.0000000   .0000000   .0000000   .0000000
 0.0000000  1.0000000   .0000000   .0000000

    18.7360050     4.1363791     1.2005194

  .0140064   .0000000   .0000000
  .0868683   .0000000   .0000000
  .2899842   .0000000   .0000000

      .3834848      .1212946

  1.0000000   .0000000
   .0000000  1.0000000

      .4600000

 1.0000000

      1.390000

 1.0000000

AL:tz2p
triple zeta plus double polarization (Ahlrichs basis)

  9
    0    0    0    0    1    1    1    2    2
    1    2    2    2    1    2    2    1    1
    5    3    2    2    5    2    2    1    1

 68090.4116770 10207.5408220  2322.8045854   656.8865384   212.9669410

  .0002841   .0000000   .0000000   .0000000   .0000000   .0000000
  .0022021   .0000000   .0000000   .0000000   .0000000   .0000000
  .0114334   .0000000   .0000000   .0000000   .0000000   .0000000
  .0462846   .0000000   .0000000   .0000000   .0000000   .0000000
  .1491057   .0000000   .0000000   .0000000   .0000000   .0000000

    75.7247166    29.0763200    11.8596239

 1.0000000   .0000000   .0000000   .0000000   .0000000
  .0000000   .4166352   .0000000   .0000000   .0000000
  .0000000   .1897837   .0000000   .0000000   .0000000

     3.3162153     1.1754789

 1.0000000   .0000000   .0000000
  .0000000  1.0000000   .0000000

      .1752053      .0647587

 1.0000000   .0000000   .0000000
  .0000000  1.0000000   .0000000

   442.7918259   104.7598884    33.3759794    12.3053880     4.8642476

  .0016391   .0000000   .0000000   .0000000   .0000000
  .0131435   .0000000   .0000000   .0000000   .0000000
  .0614977   .0000000   .0000000   .0000000   .0000000
  .1882591   .0000000   .0000000   .0000000   .0000000
  .3606261   .0000000   .0000000   .0000000   .0000000

     1.9604027      .7843673

 1.0000000   .0000000   .0000000   .0000000
  .0000000  1.0000000   .0000000   .0000000

      .1901417      .0558744

 1.0000000   .0000000
  .0000000  1.0000000

      .17000000

 1.0000000

      .52000000

 1.0000000

\end{verbatim}
 
\section{Absolute NMR shielding constants as obtained in CD based GIAO-MP2 calculations}

In this section we report the isotropic shielding constants as obtained in the representative CD based GIAO-MP2 calculations described in section IV.C of
the main document. The order of the atoms corresponds to the one in section A and the third column gives the isotropic shielding constants in ppm. \\

 \noindent
   Cholesky threshold = 4
   
coronene, dzp basis
\begin{verbatim}
       1            C               83.048           
       2            C               83.072           
       3            C               83.055           
       4            C               83.022           
       5            C               83.047           
       6            C               83.045           
       7            C               77.918      
       8            C               77.934           
       9            C               77.924           
      10            C               77.917           
      11            C               77.931           
      12            C               77.930          
      13            C               82.251          
      14            C               82.252          
      15            C               82.266           
      16            C               82.261           
      17            C               82.265           
      18            C               82.274           
      19            C               82.274           
      20            C               82.271          
      21            C               82.255           
      22            C               82.266           
      23            C               82.246           
      24            C               82.245           
      25            H               22.224             
      26            H               22.224            
      27            H               22.223            
      28            H               22.224             
      29            H               22.224             
      30            H               22.225            
      31            H               22.225             
      32            H               22.224             
      33            H               22.224            
      34            H               22.223            
      35            H               22.223            
      36            H               22.224            
\end{verbatim}
coronene, tz2p basis
\begin{verbatim}
       1            C               69.482      
       2            C               69.494           
       3            C               69.492           
       4            C               69.491          
       5            C               69.495      
       6            C               69.482           
       7            C               62.441           
       8            C               62.444          
       9            C               62.445      
      10            C               62.445           
      11            C               62.444           
      12            C               62.441          
      13            C               67.268          
      14            C               67.271      
      15            C               67.283          
      16            C               67.272      
      17            C               67.285       
      18            C               67.284          
      19            C               67.286          
      20            C               67.284           
      21            C               67.269          
      22            C               67.279           
      23            C               67.268          
      24            C               67.269           
      25            H               21.806            
      26            H               21.806             
      27            H               21.806            
      28            H               21.806            
      29            H               21.806            
      30            H               21.807             
      31            H               21.807            
      32            H               21.806             
      33            H               21.806             
      34            H               21.806           
      35            H               21.806             
      36            H               21.806             
 \end{verbatim}
 hexabenzocorone, dzp
 \begin{verbatim}
       1            C               82.139           165.417
       2            C               82.167           165.441
       3            C               82.169           165.430
       4            C               82.140           165.419
       5            C               82.164           165.444
       6            C               82.171           165.432
       7            C               78.862           168.123
       8            C               78.859           168.133
       9            C               78.851           168.135
      10            C               78.863           168.119
      11            C               78.860           168.138
      12            C               78.859           168.132
      13            C               75.081           162.568
      14            C               75.077           162.575
      15            C               75.077           162.590
      16            C               75.080           162.568
      17            C               75.073           162.585
      18            C               75.074           162.595
      19            C               75.075           162.580
      20            C               75.079           162.577
      21            C               75.080           162.592
      22            C               75.074           162.582
      23            C               75.080           162.575
      24            C               75.079           162.599
      25            C               86.862           150.581
      26            C               86.847           150.627
      27            C               86.863           150.608
      28            C               86.863           150.579
      29            C               86.846           150.632
      30            C               86.860           150.613
      31            C               86.853           150.596
      32            C               86.858           150.608
      33            C               86.854           150.620
      34            C               86.854           150.595
      35            C               86.860           150.603
      36            C               86.856           150.620
      37            C               82.123           153.762
      38            C               82.099           153.763
      39            C               82.104           153.756
      40            C               82.123           153.763
      41            C               82.102           153.758
      42            C               82.104           153.761
      43            H               21.752            10.660
      44            H               21.753            10.661
      45            H               21.753            10.659
      46            H               21.752            10.660
      47            H               21.752            10.660
      48            H               21.753            10.659
      49            H               21.752            10.659
      50            H               21.752            10.660
      51            H               21.753            10.659
      52            H               21.752            10.658
      53            H               21.753            10.660
      54            H               21.753            10.659
      55            H               22.940             5.413
      56            H               22.941             5.413
      57            H               22.942             5.412
      58            H               22.940             5.413
      59            H               22.941             5.413
      60            H               22.942             5.413
 \end{verbatim}
 
 hexabenzocoronene, tz2p
 \begin{verbatim}
       1            C               70.814           
       2            C               70.808           
       3            C               70.804          
       4            C               70.814          
       5            C               70.806          
       6            C               70.809           
       7            C               65.847          
       8            C               65.873           
       9            C               65.875          
      10            C               65.848          
      11            C               65.875           
      12            C               65.873          
      13            C               60.831          
      14            C               60.893           
      15            C               60.846          
      16            C               60.831           
      17            C               60.892           
      18            C               60.846           
      19            C               60.830           
      20            C               60.895          
      21            C               60.846           
      22            C               60.830           
      23            C               60.891      
      24            C               60.846           
      25            C               71.601           
      26            C               71.573           
      27            C               71.607           
      28            C               71.608           
      29            C               71.571           
      30            C               71.605           
      31            C               71.607           
      32            C               71.573      
      33            C               71.607           
      34            C               71.603      
      35            C               71.571           
      36            C               71.609           
      37            C               66.947           
      38            C               66.963           
      39            C               66.962           
      40            C               66.945           
      41            C               66.962           
      42            C               66.963          
      43            H               21.355            
      44            H               21.359            
      45            H               21.358            
      46            H               21.355           
      47            H               21.358           
      48            H               21.358           
      49            H               21.355            
      50            H               21.359           
      51            H               21.358            
      52            H               21.355           
      53            H               21.358            
      54            H               21.358            
      55            H               22.607             
      56            H               22.613           
      57            H               22.613            
      58            H               22.607             
      59            H               22.613           
      60            H               22.613           
 \end{verbatim}
 B$_4$tBu$_4$, dzp basis
 \begin{verbatim}
       1            B               -0.492            
       2            B               -0.498             
       3            B               -0.502             
       4            B               -0.504             
       5            C              172.620             
       6            C              160.725            
       7            C              160.740            
       8            C              160.740            
       9            H               28.437            
      10            H               30.081             
      11            H               30.081             
      12            H               28.437            
      13            H               30.081             
      14            H               30.080             
      15            H               28.437            
      16            H               30.080             
      17            H               30.081            
      18            C              172.606             
      19            C              172.615            
      20            C              172.623            
      21            C              160.744            
      22            C              160.729            
      23            C              160.731            
      24            C              160.741            
      25            C              160.732            
      26            C              160.744            
      27            C              160.726            
      28            C              160.727            
      29            C              160.726            
      30            H               28.437            
      31            H               28.437            
      32            H               28.437            
      33            H               28.437            
      34            H               28.438            
      35            H               28.438            
      36            H               28.437            
      37            H               28.437            
      38            H               28.437            
      39            H               30.081            
      40            H               30.081             
      41            H               30.081             
      42            H               30.081             
      43            H               30.081            
      44            H               30.081             
      45            H               30.081             
      46            H               30.081             
      47            H               30.081             
      48            H               30.081             
      49            H               30.081            
      50            H               30.082             
      51            H               30.080            
      52            H               30.081             
      53            H               30.081             
      54            H               30.081            
      55            H               30.081             
      56            H               30.082             
\end{verbatim}

 B$_4$tBu$_4$, tz2p basis
 \begin{verbatim}
       1            B              -15.671            
       2            B              -15.670            
       3            B              -15.669           
       4            B              -15.669            
       5            C              164.755             
       6            C              151.326            
       7            C              151.324            
       8            C              151.324            
       9            H               28.253            
      10            H               30.063             
      11            H               30.063             
      12            H               28.253            
      13            H               30.063             
      14            H               30.063           
      15            H               28.253            
      16            H               30.063            
      17            H               30.063             
      18            C              164.756            
      19            C              164.755             
      20            C              164.755           
      21            C              151.323           
      22            C              151.327           
      23            C              151.325            
      24            C              151.323           
      25            C              151.327            
      26            C              151.324            
      27            C              151.323            
      28            C              151.324            
      29            C              151.326            
      30            H               28.253            
      31            H               28.253           
      32            H               28.253           
      33            H               28.253           
      34            H               28.253            
      35            H               28.253            
      36            H               28.253            
      37            H               28.253           
      38            H               28.253            
      39            H               30.063             
      40            H               30.063             
      41            H               30.063             
      42            H               30.063             
      43            H               30.063             
      44            H               30.063             
      45            H               30.063            
      46            H               30.063             
      47            H               30.063            
      48            H               30.063             
      49            H               30.063             
      50            H               30.063            
      51            H               30.063            
      52            H               30.063             
      53            H               30.063             
      54            H               30.063             
      55            H               30.063            
      56            H               30.063             
 \end{verbatim}
Al$_4$Cp$_4$, dzp basis
\begin{verbatim}
       1            AL             716.055           
       2            AL             716.087          
       3            AL             716.061          
       4            C              100.380           
       5            C               99.726           
       6            C               99.704          
       7            C              100.380           
       8            C               99.459          
       9            H               25.517             
      10            H               25.440             
      11            H               25.441             
      12            H               25.519             
      13            H               25.404             
      14            AL             716.080           
      15            C              100.374           
      16            C              100.387           
      17            C              100.385           
      18            C              100.388           
      19            C              100.365           
      20            C              100.387           
      21            C               99.740           
      22            C               99.716           
      23            C               99.737          
      24            C               99.733           
      25            C               99.733           
      26            C               99.711           
      27            C               99.471           
      28            C               99.483           
      29            C               99.458           
      30            H               25.518             
      31            H               25.518             
      32            H               25.518             
      33            H               25.518            
      34            H               25.517             
      35            H               25.519             
      36            H               25.442             
      37            H               25.440             
      38            H               25.442             
      39            H               25.441             
      40            H               25.442            
      41            H               25.441             
      42            H               25.405             
      43            H               25.404             
      44            H               25.404             

\end{verbatim}

Al$_4$Cp$_4$, tz2p basis
\begin{verbatim}
      1            AL             716.116           
       2            AL             716.121          
       3            AL             716.122         
       4            C               85.097           
       5            C               84.391           
       6            C               84.393          
       7            C               85.091           
       8            C               84.062           
       9            H               25.344            
      10            H               25.265             
      11            H               25.265            
      12            H               25.344             
      13            H               25.208             
      14            AL             716.115           
      15            C               85.092           
      16            C               85.097           
      17            C               85.093          
      18            C               85.098          
      19            C               85.093           
      20            C               85.095           
      21            C               84.392           
      22            C               84.392           
      23            C               84.393           
      24            C               84.392           
      25            C               84.392          
      26            C               84.393           
      27            C               84.060           
      28            C               84.059           
      29            C               84.059           
      30            H               25.344             
      31            H               25.344             
      32            H               25.344             
      33            H               25.344             
      34            H               25.344             
      35            H               25.344            
      36            H               25.265             
      37            H               25.265             
      38            H               25.265             
      39            H               25.265             
      40            H               25.265             
      41            H               25.265             
      42            H               25.208             
      43            H               25.208             
      44            H               25.208            
\end{verbatim}
C$_{60}$, dzp basis
\begin{verbatim}
       1            C               66.986           
       2            C               67.000         
       3            C               66.997          
       4            C               67.003           
       5            C               67.013           
       6            C               67.005           
       7            C               66.980           
       8            C               67.025           
       9            C               66.990          
      10            C               66.989           
      11            C               67.046         
      12            C               66.998          
      13            C               67.008           
      14            C               67.016           
      15            C               67.014          
      16            C               66.971           
      17            C               67.014          
      18            C               66.977           
      19            C               67.007          
      20            C               66.966           
      21            C               67.019           
      22            C               66.980           
      23            C               66.957           
      24            C               67.008           
      25            C               66.989           
      26            C               66.975           
      27            C               67.028           
      28            C               66.978          
      29            C               67.005          
      30            C               67.045           
      31            C               66.966          
      32            C               67.007           
      33            C               66.977          
      34            C               67.014           
      35            C               66.971           
      36            C               66.989           
      37            C               67.008           
      38            C               66.957           ´
      39            C               66.980           
      40            C               67.019           
      41            C               67.045           
      42            C               67.005           
      43            C               66.978           
      44            C               67.028           
      45            C               66.975           
      46            C               66.989           
      47            C               66.990           
      48            C               67.025          ´
      49            C               66.980           
      50            C               67.005           
      51            C               67.014           
      52            C               67.016           
      53            C               67.008           
      54            C               66.998           
      55            C               67.046           
      56            C               67.003           
      57            C               66.997           
      58            C               67.000          
      59            C               66.986           
      60            C               67.013           

\end{verbatim}

C$_{60}$, tz2p basis
\begin{verbatim}
       1            C               52.363          
       2            C               52.358           
       3            C               52.369          
       4            C               52.357           
       5            C               52.364          
       6            C               52.361           
       7            C               52.363           
       8            C               52.364           
       9            C               52.367           
      10            C               52.365           
      11            C               52.364           
      12            C               52.363           
      13            C               52.363          
      14            C               52.368           
      15            C               52.363           
      16            C               52.362           
      17            C               52.367          
      18            C               52.364           
      19            C               52.360           
      20            C               52.362           
      21            C               52.366          
      22            C               52.359           
      23            C               52.363           
      24            C               52.366           
      25            C               52.361          
      26            C               52.364           
      27            C               52.361           
      28            C               52.363           
      29            C               52.368           
      30            C               52.359           
      31            C               52.362           
      32            C               52.360           
      33            C               52.364           
      34            C               52.367           
      35            C               52.362           
      36            C               52.361           
      37            C               52.366           
      38            C               52.363           
      39            C               52.359           
      40            C               52.366           
      41            C               52.359           
      42            C               52.368           
      43            C               52.363           
      44            C               52.361          
      45            C               52.364          
      46            C               52.365           
      47            C               52.367           
      48            C               52.364           
      49            C               52.363           
      50            C               52.361           
      51            C               52.363           
      52            C               52.368           
      53            C               52.363           
      54            C               52.363           
      55            C               52.364           
      56            C               52.357           
      57            C               52.369           
      58            C               52.358           
      59            C               52.363           
      60            C               52.364           ´
\end{verbatim}

 \noindent
   Cholesky threshold = 5

coronene, dzp basis
\begin{verbatim}
       1            C               83.007          
       2            C               83.012         
       3            C               83.011        
       4            C               83.010        
       5            C               83.013          
       6            C               83.007           
       7            C               77.889           
       8            C               77.899           
       9            C               77.895           
      10            C               77.896           
      11            C               77.900           
      12            C               77.888           
      13            C               82.237           
      14            C               82.241           
      15            C               82.256           
      16            C               82.241           
      17            C               82.257           
      18            C               82.256      
      19            C               82.255           
      20            C               82.257           
      21            C               82.242           
      22            C               82.254           
      23            C               82.241           
      24            C               82.237          
      25            H               22.225            
      26            H               22.225             
      27            H               22.224           
      28            H               22.225             
      29            H               22.225           
      30            H               22.225            
      31            H               22.226             
      32            H               22.225             
      33            H               22.225             
      34            H               22.224            
      35            H               22.225            
      36            H               22.225             
\end{verbatim}

coronene, tz2p basis 
\begin{verbatim}
       1            C               69.493        
       2            C               69.505          
       3            C               69.502          
       4            C               69.500          
       5            C               69.506          
       6            C               69.495           
       7            C               62.449          
       8            C               62.449           
       9            C               62.451
      10            C               62.453          
      11            C               62.451        
      12            C               62.449         
      13            C               67.274           
      14            C               67.278          
      15            C               67.290          
      16            C               67.278          
      17            C               67.293          
      18            C               67.292         
      19            C               67.292     
      20            C               67.292           
      21            C               67.278          
      22            C               67.291           
      23            C               67.278          
      24            C               67.274           
      25            H               21.806          
      26            H               21.806         
      27            H               21.806         
      28            H               21.806             
      29            H               21.806             
      30            H               21.806             
      31            H               21.806             
      32            H               21.806            
      33            H               21.806             
      34            H               21.806           
      35            H               21.806             
      36            H               21.806            
\end{verbatim}

hexabenzocorene, dzp basis
\begin{verbatim} 
       1            C               82.098        
       2            C               82.121 
       3            C               82.121 
       4            C               82.097          
       5            C               82.121         
       6            C               82.120          
       7            C               78.814           
       8            C               78.815           
       9            C               78.814          
      10            C               78.814          
      11            C               78.814           
      12            C               78.814          
      13            C               75.039           
      14            C               75.042          
      15            C               75.040           
      16            C               75.039          
      17            C               75.042           
      18            C               75.040           
      19            C               75.037           
      20            C               75.040           
      21            C               75.040           
      22            C               75.037           
      23            C               75.040           
      24            C               75.040           
      25            C               86.836           
      26            C               86.839           
      27            C               86.834           
      28            C               86.836           
      29            C               86.839          
      30            C               86.834           
      31            C               86.834           
      32            C               86.839           
      33            C               86.834           
      34            C               86.834           
      35            C               86.839           
      36            C               86.835           
      37            C               82.096           
      38            C               82.091          
      39            C               82.090           
      40            C               82.096           
      41            C               82.091           
      42            C               82.090           
      43            H               21.754            
      44            H               21.754            
      45            H               21.754            
      46            H               21.754           
      47            H               21.754            
      48            H               21.754            
      49            H               21.755            
      50            H               21.754            
      51            H               21.754            
      52            H               21.755            
      53            H               21.754            
      54            H               21.754            
      55            H               22.941             
      56            H               22.942             
      57            H               22.942             
      58            H               22.941             
      59            H               22.942             
      60            H               22.942              
\end{verbatim}

hexabenzocoronene, tz2p basis
\begin{verbatim}
       1            C               70.835      
       2            C               70.830      
       3            C               70.830     
       4            C               70.836     
       5            C               70.830     
       6            C               70.831      
       7            C               65.958      
       8            C               65.903      
       9            C               65.905     
      10            C               65.958    
      11            C               65.905    
      12            C               65.903   
      13            C               60.883   
      14            C               60.902    
      15            C               60.914    
      16            C               60.884   
      17            C               60.901    
      18            C               60.915    
      19            C               60.881   
      20            C               60.901   
      21            C               60.914   
      22            C               60.884   
      23            C               60.901    
      24            C               60.915   
      25            C               71.591   
      26            C               71.591   
      27            C               71.565   
      28            C               71.590    
      29            C               71.592    
      30            C               71.565    
      31            C               71.591     
      32            C               71.591     
      33            C               71.566    
      34            C               71.591     
      35            C               71.592    
      36            C               71.565     
      37            C               66.950       
      38            C               66.948    
      39            C               66.945   
      40            C               66.951   
      41            C               66.945  
      42            C               66.946  
      43            H               21.361   
      44            H               21.357   
      45            H               21.353   
      46            H               21.361   
      47            H               21.357   
      48            H               21.353  
      49            H               21.361 
      50            H               21.357  
      51            H               21.353   
      52            H               21.361    
      53            H               21.357    
      54            H               21.353   
      55            H               22.610    
      56            H               22.615  
      57            H               22.615  
      58            H               22.610  
      59            H               22.615   
      60            H               22.615 
\end{verbatim}

B$_4$tBu$_4$, dzp basis
\begin{verbatim}
       1            B               -0.408          
       2            B               -0.407           
       3            B               -0.409         
       4            B               -0.410         
       5            C              172.646             
       6            C              160.698           
       7            C              160.695            
       8            C              160.696           
       9            H               28.435           
      10            H               30.080           
      11            H               30.080           
      12            H               28.435          
      13            H               30.080          
      14            H               30.080           
      15            H               28.435            
      16            H               30.080            
      17            H               30.080            
      18            C              172.645             
      19            C              172.646             
      20            C              172.648             
      21            C              160.695            
      22            C              160.695           
      23            C              160.697            
      24            C              160.696            
      25            C              160.694            
      26            C              160.694            
      27            C              160.695            
      28            C              160.695            
      29            C              160.702            
      30            H               28.435            
      31            H               28.435            
      32            H               28.435            
      33            H               28.435            
      34            H               28.435            
      35            H               28.435            
      36            H               28.435            
      37            H               28.435            
      38            H               28.435            
      39            H               30.080             
      40            H               30.080             
      41            H               30.080             
      42            H               30.080             
      43            H               30.080             
      44            H               30.080             
      45            H               30.080             
      46            H               30.080             
      47            H               30.080             
      48            H               30.080             
      49            H               30.080             
      50            H               30.080             
      51            H               30.080             
      52            H               30.080             
      53            H               30.080             
      54            H               30.080             
      55            H               30.080             
      56            H               30.080             
\end{verbatim}

B$_4$tBu$_4$, tz2p basis
\begin{verbatim}
       1            B              -15.666            
       2            B              -15.667            
       3            B              -15.667            
       4            B              -15.666           
       5            C              164.755             
       6            C              151.327            
       7            C              151.326            
       8            C              151.326            
       9            H               28.253            
      10            H               30.062            
      11            H               30.063             
      12            H               28.253            
      13            H               30.062             
      14            H               30.062           
      15            H               28.253            
      16            H               30.062             
      17            H               30.062             
      18            C              164.756             
      19            C              164.755             
      20            C              164.756             
      21            C              151.325            
      22            C              151.327            
      23            C              151.325           
      24            C              151.325            
      25            C              151.326            
      26            C              151.325            
      27            C              151.326            
      28            C              151.326            
      29            C              151.326            
      30            H               28.253            
      31            H               28.253            
      32            H               28.253            
      33            H               28.253            
      34            H               28.253            
      35            H               28.253            
      36            H               28.253            
      37            H               28.253            
      38            H               28.253            
      39            H               30.062             
      40            H               30.063             
      41            H               30.062             
      42            H               30.062             
      43            H               30.062             
      44            H               30.062             
      45            H               30.062             
      46            H               30.062             
      47            H               30.062            
      48            H               30.062             
      49            H               30.062             
      50            H               30.062             
      51            H               30.062             
      52            H               30.062             
      53            H               30.062            
      54            H               30.062             
      55            H               30.062            
      56            H               30.062             
\end{verbatim}

Al$_4$Cp$_4$, dzp basis
\begin{verbatim}
       1            AL             716.314          
       2            AL             716.310          
       3            AL             716.311          
       4            C              100.399           
       5            C               99.766           
       6            C               99.764          
       7            C              100.399           
       8            C               99.529           
       9            H               25.517             
      10            H               25.442             
      11            H               25.441             
      12            H               25.518             
      13            H               25.405             
      14            AL             716.310           
      15            C              100.400          
      16            C              100.400          
      17            C              100.398           
      18            C              100.400          
      19            C              100.396          
      20            C              100.400           
      21            C               99.764           
      22            C               99.764           
      23            C               99.765           
      24            C               99.765           
      25            C               99.767           
      26            C               99.766           
      27            C               99.528           
      28            C               99.528           
      29            C               99.528           
      30            H               25.517             
      31            H               25.517             
      32            H               25.517             
      33            H               25.517             
      34            H               25.517             
      35            H               25.517             
      36            H               25.441             
      37            H               25.441             
      38            H               25.441             
      39            H               25.441             
      40            H               25.442            
      41            H               25.442             
      42            H               25.404            
      43            H               25.405             
      44            H               25.404             
  ----------------------------------------
\end{verbatim}
    
Al$_4$Cp$_4$, tz2p basis
\begin{verbatim}
       1            AL             716.162           
       2            AL             716.162           
       3            AL             716.162           
       4            C               85.102           
       5            C               84.394           
       6            C               84.393           
       7            C               85.102           
       8            C               84.048           
       9            H               25.344             
      10            H               25.265             
      11            H               25.265             
      12            H               25.344             
      13            H               25.208             
      14            AL             716.162           
      15            C               85.102           
      16            C               85.102           
      17            C               85.102           
      18            C               85.102           
      19            C               85.102           
      20            C               85.102           
      21            C               84.394           
      22            C               84.394           
      23            C               84.394           
      24            C               84.393           
      25            C               84.393           
      26            C               84.393           
      27            C               84.048           
      28            C               84.048           
      29            C               84.048           
      30            H               25.344             
      31            H               25.344             
      32            H               25.344             
      33            H               25.344             
      34            H               25.344             
      35            H               25.344             
      36            H               25.265             
      37            H               25.265             
      38            H               25.265             
      39            H               25.265             
      40            H               25.265             
      41            H               25.265             
      42            H               25.208             
      43            H               25.208             
      44            H               25.208             
\end{verbatim}

C$_{60}$, dzp basis
\begin{verbatim}
       1            C               67.007           
       2            C               67.040           
       3            C               67.015          
       4            C               67.040         
       5            C               67.010         
       6            C               67.008          
       7            C               67.010          
       8            C               67.039        
       9            C               67.013           
      10            C               67.038           
      11            C               67.040           
      12            C               67.015         
      13            C               67.042          
      14            C               67.009           
      15            C               67.009         
      16            C               67.011           
      17            C               67.040           
      18            C               67.014          
      19            C               67.039          
      20            C               67.013         
      21            C               67.037           
      22            C               67.010           
      23            C               67.011          
      24            C               67.037           
      25            C               67.013         
      26            C               67.016       
      27            C               67.042           
      28            C               67.011          
      29            C               67.012           
      30            C               67.043           
      31            C               67.013           
      32            C               67.039         
      33            C               67.014           
      34            C               67.040         
      35            C               67.011           
      36            C               67.013           
      37            C               67.037           
      38            C               67.011         
      39            C               67.010           
      40            C               67.037           
      41            C               67.043          
      42            C               67.012           
      43            C               67.011           
      44            C               67.042          
      45            C               67.016          
      46            C               67.038           
      47            C               67.013           
      48            C               67.039         
      49            C               67.010          
      50            C               67.008           
      51            C               67.009           
      52            C               67.009         
      53            C               67.042          
      54            C               67.015           
      55            C               67.040        
      56            C               67.040           
      57            C               67.015     
      58            C               67.040       
      59            C               67.007     
      60            C               67.010           
\end{verbatim}

C$_{60}$, tz2p basis
\begin{verbatim}
       1            C               52.374           
       2            C               52.377           
       3            C               52.376          
       4            C               52.376           
       5            C               52.376           
       6            C               52.375           
       7            C               52.375           
       8            C               52.376           
       9            C               52.377           
      10            C               52.375           
      11            C               52.376          
      12            C               52.374          
      13            C               52.376         
      14            C               52.374           
      15            C               52.376          
      16            C               52.375           
      17            C               52.376          
      18            C               52.375           
      19            C               52.376           
      20            C               52.376           
      21            C               52.376           
      22            C               52.376         
      23            C               52.375           
      24            C               52.377         
      25            C               52.374           
      26            C               52.374           
      27            C               52.376           
      28            C               52.375          
      29            C               52.374           
      30            C               52.376          
      31            C               52.376           
      32            C               52.376           
      33            C               52.375           
      34            C               52.376          
      35            C               52.375         
      36            C               52.374           
      37            C               52.377           
      38            C               52.375           
      39            C               52.376           
      40            C               52.376           
      41            C               52.376           
      42            C               52.374           
      43            C               52.375          
      44            C               52.376           
      45            C               52.374           
      46            C               52.375           
      47            C               52.377           
      48            C               52.376           
      49            C               52.375          
      50            C               52.375           
      51            C               52.376           
      52            C               52.374           
      53            C               52.376          
      54            C               52.374           
      55            C               52.376          
      56            C               52.376         
      57            C               52.376           
      58            C               52.377          
      59            C               52.374           
      60            C               52.376          
 \end{verbatim}

tweezer host-guest complex, dzp basis
\begin{verbatim}
       1            N               49.894           
       2            H               23.996           
       3            H               27.124             
       4            H               23.447            
       5            C               86.929           
       6            H               24.427             
       7            H               27.279            
       8            H               23.903            
       9            H               26.003            
      10            H               27.252            
      11            H               24.251             
      12            H               27.352             
      13            H               24.397             
      14            C               58.771           
      15            C               87.988           
      16            H               24.427            
      17            C               84.977           
      18            H               25.079             
      19            C               83.982           
      20            H               24.993           
      21            C               85.553           
      22            C               58.190           
      23            C              148.659            
      24            C              148.485            
      25            H               27.279            
      26            C              134.767            
      27            H               28.700            
      28            H               28.800            
      29            C               60.221           
      30            C               92.735           
      31            H               23.903             
      32            C               61.465      
      33            C               61.124           
      34            C               88.516           
      35            C               59.601           
      36            C               81.626          
      37            H               26.003             
      38            C               81.039           
      39            H               29.676             
      40            C               93.772           
      41            N               49.897          
      42            C               58.773          
      43            C               87.988           
      44            C               84.976           
      45            H               25.080             
      46            C               83.985      
      47            H               24.994            
      48            C               85.554           
      49            H               23.996            
      50            C               58.190           
      51            C              148.660            
      52            H               27.125             
      53            C              148.485             
      54            C               60.216           
      55            C               92.733           
      56            C               61.467           
      57            C               61.125           
      58            C               88.516           
      59            H               23.447             
      60            C               59.600      
      61            C               81.626           
      62            C               81.040           
      63            H               29.678            
      64            C               93.768           
      65            C               86.926           
      66            C              134.771            
      67            H               28.700            
      68            H               28.800            
      69            C              149.182             
      70            C              149.014             
      71            H               27.352             
      72            C              136.144            
      73            H               29.002            
      74            H               28.774            
      75            C               62.172          
      76            C               88.456           
      77            H               24.397             
      78            C               77.997           
      79            C               87.483          
      80            C               61.815           
      81            C              149.183             
      82            H               27.252             
      83            C              149.013             
      84            C              136.145            
      85            H               29.002            
      86            H               28.774           
      87            C               62.169           
      88            C               88.454           
      89            C               78.000           
      90            C               87.482           
      91            H               24.251             
      92            C               61.815           
\end{verbatim}

tweezer host-guest complex, tzp basis
\begin{verbatim}
       1            N               23.212           
       2            H               24.068             
       3            H               27.164             
       4            H               23.571             
       5            C               67.826           
       6            H               24.531            
       7            H               27.329             
       8            H               24.067             
       9            H               26.182             
      10            H               27.279            
      11            H               24.338            
      12            H               27.382             
      13            H               24.495             
      14            C               43.532           
      15            C               74.049           
      16            H               24.531             
      17            C               70.758           
      18            H               25.160             
      19            C               70.141           
      20            H               25.071             
      21            C               71.184           
      22            C               43.009           
      23            C              139.656             
      24            C              139.272             
      25            H               27.329             
      26            C              125.153            
      27            H               28.815            
      28            H               28.907            
      29            C               45.173           
      30            C               78.893           
      31            H               24.067             
      32            C               46.521           
      33            C               46.538           
      34            C               73.932           
      35            C               44.810          
      36            C               67.323           
      37            H               26.182            
      38            C               67.056           
      39            H               29.965             
      40            C               81.818           
      41            N               23.213           
      42            C               43.531           
      43            C               74.050           
      44            C               70.759          
      45            H               25.161            
      46            C               70.143           
      47            H               25.072           
      48            C               71.185           
      49            H               24.068           
      50            C               43.008           
      51            C              139.657            
      52            H               27.164            
      53            C              139.272            
      54            C               45.169      
      55            C               78.892           
      56            C               46.521           
      57            C               46.537           
      58            C               73.932           
      59            H               23.571            
      60            C               44.811           
      61            C               67.319      
      62            C               67.056          
      63            H               29.966           
      64            C               81.814           
      65            C               67.827           
      66            C              125.158            
      67            H               28.815            
      68            H               28.907            
      69            C              140.231             
      70            C              139.953             
      71            H               27.382             
      72            C              126.637            
      73            H               29.097            
      74            H               28.877            
      75            C               47.638           
      76            C               75.639           
      77            H               24.495             
      78            C               64.906           
      79            C               74.148           
      80            C               47.323           
      81            C              140.231             
      82            H               27.278             
      83            C              139.954             
      84            C              126.639            
      85            H               29.097            
      86            H               28.877            
      87            C               47.637           
      88            C               75.638           
      89            C               64.907           
      90            C               74.148           
      91            H               24.338             
      92            C               47.323           
\end{verbatim}